\documentclass{article}

\usepackage{amssymb}
\usepackage{amsmath}
\usepackage{epsfig}
\begin{document}

\title{\bf Cold Plasma Dispersion Relations in the Vicinity of a Schwarzschild Black Hole Horizon}

\author{M. Sharif \thanks{msharif@math.pu.edu.pk} and Umber Sheikh\\
\textit{\ {\small Department of Mathematics,}}\\
\textit{{\small University of the Punjab, Lahore 54590,
Pakistan}}}
\date{}
\maketitle
\begin{abstract}
We apply the ADM 3+1 formalism to derive the general relativistic
magnetohydrodynamic equations for cold plasma in spatially flat
Schwarzschild metric. Respective perturbed equations are
linearized for non-magnetized and magnetized plasmas both in
non-rotating and rotating backgrounds. These are then Fourier
analyzed and the corresponding dispersion relations are obtained.
These relations are discussed for the existence of waves with
positive angular frequency in the region near the horizon. Our
results support the fact that no information can be extracted from
the Schwarzschild black hole. It is concluded that negative phase
velocity propagates in the rotating background whether the black
hole is rotating or non-rotating.
\end{abstract}


\section{Introduction}

General Relativity (GR) serves both as a theory of space and time
as well as a theory of gravitation. This theory has tremendous
philosophical implications and has given rise to exotic new
physical concepts like black holes and dark matter. All the
predictions and applications by different scientists make the
theory a spectacularly successful one. The theory of GR is not
complete unless it includes the hard reality of experiments and of
astronomical observations. To explain the phenomena observed by
astronomers in physical terms we have astrophysical relativity.
This serves as a tool to gauge the properties of large scale
structures for which gravitation plays a significant role.

It is believed that the compact objects have strong gravitational
fields near their surfaces [1]. The study of general relativistic
effects on electromagnetic processes which take place in the
vicinity of compact objects is of theoretical interest in its own
right. However, it is also important to interpret and understand
the still puzzling signals we receive from these astronomical
objects, including neutron stars and black holes. All the massive
stars, neutron stars and black holes carry energy flux. Due to
this flux, a relatively large magnetic field is produced. In order
to understand the phenomena of the compact objects and many more
like these, we have a developed theory of magnetized plasma called
theory of magnetohydrodynamics (MHD). To study the magnetosphere
of massive black holes, the strength of gravity demands the theory
of general relativistic magnetohydrodynamics (GRMHD). GRMHD
equations help us to study stationary configurations and dynamic
evolution of conducting fluid in a magnetosphere. These equations
(Maxwell's equations, Ohm's law, mass, momentum and energy
conservation equations) are required to investigate various
aspects of the interaction of relativistic gravity with plasma's
magnetic field. The successful study of plasmas in the black hole
environment is important because a successful study of waves will
be of great value in aiding the observational identification of
black hole candidates. It is well-known [2-4] that an isolated
black hole cannot have an electromagnetic field unless it is
endowed with a net electric charge. Since a collapsed object can
have a very strong effect on an electromagnetic field, it is of
interest to determine this effect when a black hole is placed in
an external electromagnetic field using GRMHD equations.

The 3+1 formalism (also called ADM formalism) was originally
developed to study the quantization of gravitational field by
Arnowitt et al. [5]. Tarfulea [6] worked on numerical solutions of
Einstein field equations by applying constraint preserving
boundary conditions. Thorne and Macdonald [7,8] extended the
formulation to electromagnetic fields. Thorne et al. [9] described
its further applications for the black hole theory. The black hole
theory was considered by Holcomb and Tajima [10], Holcomb [11] and
Dettmann et al. [12] to investigate some properties of wave
propagation in Friedmann universe. Khanna [13] derived the MHD
equations describing the two component plasma theory of Kerr black
hole in 3+1 split. Ant\'{o}n et al. [14] used 3+1 formalism to
investigate various test simulations and discussed
magneto-rotational instability of accretion disks. Anile [15]
worked on propagation and stability of relativistic shocks and
relativistic simple waves in magneto-fluids. He also linearized
the electromagnetic wave theory in cold relativistic plasma.
Komissarov [16] discussed Blendford-Znajec monopole solution using
3+1 formalism in black hole electrodynamics. In another paper
[17], he formulated and solved time dependent, force free,
degenerate electrodynamics as a hyperbolic system of conservations
laws for pulsars. He also discussed the waves mode and one
dimensional numerical scheme based on linear and exact Riemann
solvers. Zhang [18] formulated the black hole theory for
stationary symmetric GRMHD using $3+1$ formalism. The same author
[19] applied this formulation to the linearized waves propagating
in two dimensions using perfect GRMHD fluid for the cold plasma
state in the vicinity of Kerr black hole magnetosphere. He
investigated the behavior of perturbations in graphic form which
described the response of the magnetosphere to oscillatory driving
forces near the plasma-injection plane. Tsikarishvili et al. [20]
considered 3+1 general relativistic hydrodynamical equations
describing strongly magnetized collisionless plasma with an
anisotropic pressure and obtained the equations of state in
ultra-relativistic limit. In his recent paper, Mofiz [21]
discussed gravitational waves in magnetized plasmas.

The formalism for gravitational perturbations away from a
Scwarzschild background was developed by Regge and Wheeler [22].
It was extended by Zerilli [23] who proved that the perturbations
corresponding to a change in mass, the angular momentum and
charge, of the Schwarzschild black hole are well-behaved. The
decay of non well-behaved perturbations was investigated by Price
[24]. The quasi-static electric problem was solved by Hanni and
Ruffini [25] who showed that the lines of force diverge at the
horizon for the observer at infinity. Wald [26] derived the
solution for electromagnetic field occurring when a stationary
axisymmetric black hole is placed in a uniform magnetic field
aligned along the symmetry axis of black hole. A linearized
treatment of plasma waves for special relativistic formulation of
Schwarzschild black holes was developed by Sakai and Kawata [27].
This treatment was extended to waves in general relativistic two
component plasma in 3+1 ADM formalism, propagating in radial
direction by Buzzi et al. [28]. They investigated the one
dimensional radial propagation of transverse and longitudinal
waves close to the Schwarzschild horizon.

This paper is devoted to study the dynamical magnetosphere of the
Schwarzschild black hole. The model spacetime is the Schwarzschild
metric in Rindler coordinates which preserves the key features of
the Schwarzschild geometry. This investigation supports the
well-known fact that the information from the black hole cannot be
extracted. We apply 3+1 form of the GRMHD equations to this
spacetime by using the cold plasma state. Afterwards, we linearize
the equations to study perturbations of the magnetosphere. These
equations are then Fourier analyzed to get the sets of ordinary
differential equations which are comparatively easier to handle
than the original GRMHD partial differential equations. The
determinant of these equations are solved [29] to get dispersion
relations. Since the determinant of the coefficients of the
Fourier analyzed equations cannot be solved analytically, we solve
it with the help of software \emph{Mathematica} for special cases
and wave numbers are obtained in terms of angular frequency and
lapse function. We use wave number to calculate phase and group
velocities, refractive index and change in refractive index w.r.t.
angular frequency. The graphs of each of these quantities are
given and discussed.

The paper has been organized as follows. In the next section, we
shall review the general results for the 3+1 formalism. The GRMHD
equations in the Schwarzschild black hole magnetosphere for the
cold plasma are also given. In section $3$, we restrict the
problem only to the non-rotating background. Section $4$ is
devoted to the rotating non-magnetized plasma and section $5$ is
furnished with the dispersion relations for the rotating
magnetized plasma. In the last section, we shall summarize and
discuss the results.

\section{3+1 Spacetime Modelling}

In the $3+1$ formalism, the line element of the spacetime can be
written as [19]
\begin{equation}
ds^2=-\alpha^2dt^2+\gamma_{ij}(dx^i+\beta^idt)(dx^j+\beta^jdt),
\end{equation}
where $\alpha$ (lapse function), $\beta^i$ (shift vector) and
$\gamma_{ij}$ (spatial metric) are functions of the coordinates
$t,~x^i$. A natural observer, associated with this spacetime
called fiducial observer (FIDO), has four velocity \textbf{n}
perpendicular to the hypersurfaces of constant time $t$ and is
given by
\begin{equation}
\bf{n}=\frac{1}{\alpha}(\frac{\partial}{\partial
t}-\beta^i\frac{\partial}{\partial x^i}).
\end{equation}
Throughout the paper, geometrized units with $G=c=1$ will be used.
Vectors and tensors living in four-dimensional spacetime will be
denoted by boldface italic letters such as FIDO's four-velocity
\textbf{n}. The three-dimensional tensors are distinguished from
vectors by a dyad over the letter, such as three-dimensional
metric $\overleftrightarrow{\gamma}$. All vector analysis
notations such as gradient, curl and vector product will be those
of the three dimensional absolute space with three metric
$\overleftrightarrow{\gamma}$. The determinant of three metric is
denoted as $g$: $g\equiv det|\gamma_{ij}|.$ Latin letters $i, j,
k,...$ represent indices in absolute space and run from 1 to 3.

The Schwarzschild line element in Rindler coordinates [9]
\begin{equation}
ds^2=-\alpha^2(z)dt^2+dx^2+dy^2+dz^2
\end{equation}
is the best approximation of this spacetime in Cartesian
coordinates. Since the Schwarzschild black hole is non-rotating,
hence the shift vector $\beta=0$. This model is an analogue of the
Schwarzschild spacetime with $z$, $x$ and $y$ as radial $r$, axial
$\phi$ and poloidal $\theta$ directions respectively. $\alpha$ is
the lapse function which depends on $r$ in the Shwarzschild
metric, whereas it depends on $z$ in this analogue. In Rindler
coordinates, $\alpha=\frac{z}{2r_H}$, where $r_H$ is the radius of
the Schwarzschild black hole. This function vanishes at the
horizon which we can place at $z=0$ and it increases monotonically
as $z$ increases from $0$ to $\infty.$ The associated FIDO's
four-velocity vector reduces to
$\textbf{n}=\frac{1}{\alpha}\frac{\partial}{\partial t}$. The
constant time surfaces have spatial geometry $ds^2=dx^2+dy^2+dz^2$
which gives a good approximation to the Schwarzschild geometry
near the horizon.

\subsection{The 3+1 Split: Laws of Physics}

The real power of 3+1 viewpoint lies in its strong similarity to a
physicists with non-general relativistic experience. The 3+1 laws
of electrodynamics are formulated with an eye towards emphasizing
this similarity. However, there are real unavoidable and crucial
new features of these familiar laws that enter electrodynamics in
the vicinity of a black hole. The most important of these is the
issue of two times, the universal time $t$ and the FIDO's time
$\tau$. Both times are related to each other by
$\alpha\equiv\frac{d\tau}{dt}$. For the Schwarzschild black hole,
the Maxwell equations in 3+1 formalism take the form [9]
\begin{eqnarray}
\nabla.\textbf{B}&=&0,\\
\frac{\partial\textbf{B}}{\partial
t}&=&-\nabla\times(\alpha\textbf{E}),\\
\nabla.\textbf{E}&=&4\pi\rho_e,\\
\frac{\partial\textbf{E}}{\partial
t}&=&\nabla\times(\alpha\textbf{B})-4\pi \alpha \textbf{j},
\end{eqnarray}
where $\textbf{B}$ and $\textbf{E}$ are the FIDO measured magnetic
and electric fields, $\textbf{j}$ and $\rho_e$ are electric
current and electric charge density respectively. Under the
perfect MHD condition, i.e.,
\begin{equation}
\textbf{E}+\textbf{V}\times\textbf{B}=0,
\end{equation}
where {\textbf V} is the velocity of the fluid measured by the
FIDO, there can be no electric field in fluid's rest frame. For
perfect MHD, the equations of evolution of magnetic field [18]
become
\begin{equation}
\frac{D\textbf{B}}{D\tau}+\frac{1}{\alpha}(\textbf{B}.\nabla)(\beta-\alpha\textbf{V})
+\left\{\theta+\frac{1}{\alpha}\nabla.(\alpha\textbf{V})\right\}\textbf{B}=0,
\end{equation}
where
\begin{equation*}
\frac{D}{D\tau}\equiv\frac{d}{d\tau}+\textbf{V}.\nabla
=\frac{1}{\alpha}\left\{\frac{\partial}{\partial
t}+(\alpha\textbf{V}-\beta).\nabla\right\}.
\end{equation*}
The expansion rate of FIDO's four-velocity is
\begin{equation*}
\theta=\frac{1}{\alpha}(\frac{\dot{g}}{2g}-\nabla.\beta).
\end{equation*}
Here dot indicates partial derivative with respect to time $t$.
The FIDO's measured rate of change of any three dimensional vector
in absolute space (orthogonal to \textbf{n}) is
\begin{equation*}
\frac{d}{d \tau}\equiv \frac{1}{\alpha}(\frac{\partial}{\partial t
}-\beta.\nabla).
\end{equation*}
Local conservation law of rest-mass according to FIDO is given by
[18]
\begin{equation*}
\frac{D\rho_0}{D\tau}+\rho_0\gamma^2\textbf{V}.\frac{D\textbf{V}}{D\tau}+
\frac{\rho_0}{\alpha}\left\{\frac{\dot{g}}{2g}+\nabla.(\alpha\textbf{V}-\beta)\right\}=0,
\end{equation*}
or
\begin{equation}
\frac{D\rho_0}{D\tau}+\rho_0\gamma^2\textbf{V}.\frac{D\textbf{V}}{D\tau}+
\rho_0\left\{\theta+\frac{1}{\alpha}\nabla.(\alpha\textbf{V})\right\}=0
\end{equation}
with $\rho_0$ as the rest mass density. FIDO measured law of force
balance equation can be written as
\begin{eqnarray}
&&\left\{\left(\rho_0\gamma^2\mu+\frac{\textbf{B}^2}{4\pi}\right)\gamma_{ij}
+\rho_0\gamma^4\mu
V_iV_j-\frac{1}{4\pi}B_iB_j\right\}\frac{DV^j}{D\tau}
+\rho_0\gamma^2V_i\frac{D\mu}{D\tau}
-\left(\frac{\textbf{B}^2}{4\pi}\gamma_{ij}
-\frac{1}{4\pi}B_iB_j\right){V^j}_{|k}V^k\nonumber\\
&& =-\rho_0\gamma^2\mu \left\{a_i-\frac{1}{\alpha}\beta_{j|i}V^j
-(\pounds_t\gamma_{ij})V^j\right\}-p_{|i}+\frac{1}{4\pi}(\textbf{V}\times
\textbf{B})_i \nabla.(\textbf{V}\times \textbf{B})
-\frac{1}{8\pi\alpha^2}(\alpha
\textbf{B})^2_{|i}\nonumber\\
&&+\frac{1}{4\pi\alpha}(\alpha
B_i)_{|j}B^j-\frac{1}{4\pi\alpha}(\textbf{B}\times\{\textbf{V}\times
[\nabla \times(\alpha \textbf{V} \times
\textbf{B})-(\textbf{B}.\nabla ) \beta]+(\textbf{V} \times
\textbf{B}). \nabla \beta \})_i
\end{eqnarray}
with $p$, the pressure. Notice that the acceleration,
$\textbf{a}=\frac{\nabla \alpha}{\alpha}$, is a function of time
lapse in 3+1 formalism.

For the Schwarzschild black hole with perfect MHD assumption given
by Eq.(2.8), the equations of evolution of magnetic field (2.5)
and (2.9), the mass  conservation law (2.10) and the momentum
conservation law (2.11) take the following forms respectively
\begin{eqnarray}
&&\frac{\partial \textbf{B}}{\partial t}=-\nabla
\times(\alpha \textbf{V}\times \textbf{B}),\\
&&\frac{\partial \textbf{B}}{\partial t}+(\alpha
\textbf{V}.\nabla)\textbf{B}-(\textbf{B}.\nabla)(\alpha \textbf{V}
)+(\nabla.(\alpha \textbf{V}))\textbf{B}=0,\\
&&\frac{\partial (\rho_0 \mu)}{\partial t}+\{(\alpha
\textbf{V}).\nabla\}(\rho_0 \mu)+\rho_0 \mu
\gamma^2\textbf{V}.\frac{\partial \textbf{V}}{\partial t}+\rho_0
\mu \gamma^2\textbf{V}.(\alpha \textbf{V}.\nabla)\textbf{V}+
\rho_0 \mu \{\nabla.(\alpha\textbf{V})\}=0,\\
&&\left\{\left(\rho_0\mu
\gamma^2+\frac{\textbf{B}^2}{4\pi}\right)\delta_{ij}+\rho_0\mu
\gamma^4
V_iV_j-\frac{1}{4\pi}B_iB_j\right\}\left(\frac{1}{\alpha}\frac{\partial}{\partial
t}+\textbf{V}.\nabla\right)V^j-\left(\frac{\textbf{B}^2}{4\pi}\delta_{ij}
-\frac{1}{4\pi}B_iB_j\right){V^j}_{,k}V^k\nonumber\\
&&+\rho_0 \gamma^2 V_i\left\{\frac{1}{\alpha}\frac{\partial
\mu}{\partial t}+(\textbf{V}.\nabla)\mu\right\}=-\rho_0\mu
\gamma^2 a_i-p_{,i}+ \frac{1}{4\pi}(\textbf{V}\times \textbf{B})_i
\nabla.(\textbf{V}\times \textbf{B})
-\frac{1}{8\pi\alpha^2}(\alpha \textbf{B})^2_{,i}\nonumber\\
&&+\frac{1}{4\pi\alpha}(\alpha
B_i)_{,j}B^j-\frac{1}{4\pi\alpha}[\textbf{B}\times\{\textbf{V}
\times (\nabla \times (\alpha \textbf{V} \times \textbf{B}))\}]_i.
\end{eqnarray}
Eqs.(2.12)-(2.15) are the perfect GRMHD equations for the
Schwarzschild black hole. In the rest of the paper, we shall
analyze these equations using perturbation and Fourier analysis
procedures.

\subsection{Cold Plasma and GRMHD Equations}

Theoretical modelling of the moving plasmas neglects the thermal
and gravity effects (such as the pressure forces acting on plasma
may not be as important as electromagnetic and centrifugal
forces). This plasma is called cold relativistic plasma. This is
the simplest closed system and contains only the equations of
conservation of mass and momentum. The highest moment of this
system, i.e., the kinematic pressure dyad is taken to be zero.
This model can be used in the study of small amplitude
electromagnetic waves propagating in plasmas with phase velocity
much larger than the thermal velocity of the particles.

In hydrodynamic treatment, the plasma is represented as a perfect
fluid which has no viscosity and no heat conduction. In Rindler
analogue of the Schwarzschild spacetime, we assume that the system
of equations for perfect MHD is enclosed by the cold plasma. This
state can be expressed by the following equation [19]
\begin{equation}
\mu=\frac{\rho}{\rho_0}=constant.
\end{equation}
We note that the cold plasma has vanishing thermal pressure and
vanishing thermal energy. Here $\rho$ is the mass density of the
fluid.

Using Eq.(2.16), the GRMHD Eqs.(2.12)-(2.15) become
\begin{eqnarray}
&&\frac{\partial \textbf{B}}{\partial t}=\nabla \times(\alpha
\textbf{V} \times \textbf{B}),\\
&&\frac{\partial \textbf{B}}{\partial t}+(\alpha
\textbf{V}.\nabla)\textbf{B}-(\textbf{B}.\nabla)(\alpha
\textbf{V})+\textbf{B}\nabla.(\alpha \textbf{V})=0,\\
&&\frac{\partial \rho}{\partial t}+(\alpha \textbf{V}.\nabla)\rho
+\rho\gamma^2\textbf{V}.\frac{\partial \textbf{V}}{\partial
t}+\rho\gamma^2\textbf{V}.(\alpha
\textbf{V}.\nabla)\textbf{V}+ \rho \nabla.(\alpha\textbf{V})=0,\\
&&\left\{\left(\rho \gamma^2+\frac{\textbf{B}^2}{4
\pi}\right)\delta_{ij} +\rho
\gamma^4V_iV_j-\frac{1}{4\pi}B_iB_j\right\}\left(\frac{1}{\alpha}\frac{\partial}{\partial
t}+\textbf{V}.\nabla
\right)V^j-\left(\frac{\textbf{B}^2}{4\pi}\delta_{ij}-\frac{1}{4
\pi}B_i B_j\right)V^j_{,k} V^k+\rho \gamma^2 a_i  \nonumber\\
&&\frac{1}{4 \pi} (\textbf{V}\times \textbf{B})_i
\nabla.(\textbf{V}\times \textbf{B}) -\frac{(\alpha
\textbf{B})^2_{,i}}{8 \pi \alpha^2} +\frac{(\alpha B_i)_{,j}
B^j}{4 \pi \alpha}-\frac{1}{4 \pi
\alpha}[\textbf{B}\times\{\textbf{V} \times (\nabla \times (\alpha
\textbf{V} \times \textbf{B}))\}]_i.
\end{eqnarray}
The perturbed flow in the magnetosphere shall be characterized by
its velocity \textbf{V}, magnetic field \textbf{B} (as measured by
the FIDO) and fluid density $\rho$. The first order perturbations
in the above mentioned quantities are denoted by
$\delta\textbf{V},~\delta \textbf{B}$ and $\delta \rho$.
Consequently, the perturbed variables will take the following
form:
\begin{equation}
\textbf{B}=\textbf{B}^0+B\textbf{b},~
\textbf{V}=\textbf{V}^0+\textbf{v},~ \rho=\rho^0+\rho
\tilde{\rho},
\end{equation}
where $\textbf{B}^0,~\textbf{V}^0$ and $\rho^0$ are unperturbed
quantities. Since waves can propagate in $z$-direction due to
gravitation with respect to time $t$, the perturbed quantities
depend on $z$ and $t$.

\section{Non-Rotating Background}

In this section, we shall solve GRMHD equations by taking
non-rotating background. Notice that this makes no difference
whether we take magnetic field zero or non-zero. The relative
assumptions for this background are given. The perturbations and
Fourier analyze techniques are used to reduce GRMHD equations to
ordinary differential equations. The numerical solutions of
dispersion relations are given in the form of graphs.

\subsection{Relative Assumptions}

For non-rotating background, the magnetosphere has the perturbed
flow only along $z$-axis. The fluid four-velocity measured by FIDO
is described by a spatial vector field lying along the $z$-axis
and is given by
\begin{equation}{\setcounter{equation}{1}}
\textbf{V}=u(z)\textbf{e}_\textbf{z}.
\end{equation}
The Lorentz factor $\gamma=\frac{1}{\sqrt{1-\textbf{V}^2}}$ which
takes the form
\begin{equation}
\gamma=\frac{1}{\sqrt{1-u^2}}.
\end{equation}
FIDO measured magnetic field is given by
\begin{equation}
\textbf{B}=B(z)\textbf{e}_\textbf{z}.
\end{equation}
The following notations for the perturbed quantities will be used
\begin{eqnarray}
\textbf{b}&\equiv& \frac{ \delta \textbf{B}}{B}
=\textbf{b}_\textbf{z}(t,z)\textbf{e}_\textbf{z},\nonumber\\
\textbf{v}&\equiv& \delta \textbf{V}=\textbf{v}
_\textbf{z}(t,z)\textbf{e}_\textbf{z},\nonumber\\
\tilde{\rho}&\equiv&\frac{\delta \rho}{\rho}=\tilde{\rho}(t,z).
\end{eqnarray}
We also assume that the perturbations have harmonic space and time
dependence, i.e.,
\begin{equation}\tilde{\rho},~\textbf{b},~
\textbf{v}\sim e^{-i (\omega t-k z)}.
\end{equation}
Thus the perturbed variables can be expressed as follows:
\begin{eqnarray}
\tilde{\rho}(t,z)=c_1e^{-\iota (\omega t-kz)},\nonumber\\
\textbf{v}_\textbf{z}(t,z)=c_2e^{-\iota (\omega t-kz)},\nonumber\\
\textbf{b}_\textbf{z}(t,z)=c_3e^{-\iota (\omega t-kz)},
\end{eqnarray}
where $c_1,~c_2,~c_3$ are arbitrary constants.

\subsection{Perturbation Equations}

We use linear perturbation to write down the GRMHD
Eqs.(2.17)-(2.20) by using Eq.(2.21)
\begin{eqnarray}
&&\frac{\partial (\delta \textbf{B})}{\partial t}=\nabla
\times(\alpha \textbf{v} \times \textbf{B})
+\nabla \times (\alpha \textbf{V} \times \delta \textbf{B}),\\
&&\nabla. (\delta \textbf{B})=0,\\
&&(\frac{1}{\alpha}\frac{\partial}{\partial t}+\textbf{V}.\nabla
)\delta \rho + \rho \gamma^2 \textbf{V}.(\frac{1}{\alpha}
\frac{\partial}{\partial t}+ \textbf{V}.\nabla)\textbf{v} -
\frac{\delta \rho}{\rho}(\textbf{V}.\nabla)\rho +\rho (\nabla.\textbf{v})\nonumber\\
&&=-2\rho \gamma^2 (\textbf{V}.\textbf{v})( \textbf{V}. \nabla) ln
\gamma -\rho \gamma^2 (\textbf{V}.\nabla \textbf{V}).\textbf{v}
+\rho (\textbf{v}.\nabla ln u),\\
&&\{(\rho\gamma^2+\frac{\textbf{B}^2}{4
\pi})\delta_{ij}+\rho\gamma^4 V_i V_j-\frac{1}{4 \pi}B_i
B_j\}\frac{1}{\alpha} \frac{\partial v^j}{\partial t} +\frac{1}{4
\pi}[\textbf{B} \times \{\textbf{V} \times
\frac{1}{\alpha}\frac{\partial(\delta
\textbf{B})}{\partial t}\}]_i\nonumber\\
&&+\rho \gamma^2v_{i,j}V^j +\rho \gamma^4 V_i v_{j,k} V^j V^k
-\frac{1}{4 \pi \alpha}\{(\alpha
\delta B_i)_{,j}-(\alpha \delta B_j)_{,i}\} B^j\nonumber\\
&&=-\gamma^2\{\delta \rho+2 \rho \gamma^2
(\textbf{V}.\textbf{v})\}a_i +\frac{1}{4 \pi \alpha}\{(\alpha
B_i)_{,j}-(\alpha B_j)_{,i}\}\delta B^j \nonumber\\
&&-\rho\gamma^4 (v_i V^j + v^j V_i)V_{k,j}V^k- \gamma^2\{\delta
\rho V^j +2 \rho \gamma^2 (\textbf{V}.\textbf{v}) V^j +\rho v^j\} V_{i,j}\nonumber\\
&&-\gamma^4 V_i\{\delta \rho V^j+4 \rho \gamma^2
(\textbf{V}.\textbf{v})V^j+\rho v^j\}V_{j,k}V^k.
\end{eqnarray}
It is to be noted that the law of conservation of rest-mass [18]
in three dimensional hypersurface $$ \alpha \rho \gamma
u=constant$$ is used to obtain Eq.(3.9). This will also be used in
deriving the component form of these equations.

The component form of Eqs.(3.7)-(3.10) is given as follows:
\begin{eqnarray}
&&\frac{1}{\alpha}\frac{\partial b_z}{\partial t}=0,\\
&&b_{z,z}=0,\\
&&\frac{1}{\alpha}\frac{\partial \tilde{\rho}}{\partial t}+u
\tilde{\rho}_{,z} +\gamma^2 u\frac{1}{\alpha} \frac{\partial
v_z}{\partial t}+(1+\gamma^2 u^2)v_{z,z}
=(1-2\gamma^2u^2)(1+\gamma^2 u^2)\frac{u'}{u}v_z,\\
&&\gamma^2(1+\gamma^2 u^2)\frac{1}{\alpha} \frac{\partial
v_z}{\partial t}+\gamma^2(1+\gamma^2 u^2)u v_{z,z}
=-\tilde{\rho}\gamma^2\{a_z+u(1+\gamma^2u^2)u'\}\nonumber\\
&&-\gamma^2\{u'(1+\gamma^2 u^2)(1+4\gamma^2 u^2)
+2u\gamma^2a_z)\}v_z.
\end{eqnarray}
When the above equations are Fourier analyzed by substituting
values from Eq.(3.6), these take the form
\begin{eqnarray}
&&-\frac{\iota \omega}{\alpha} c_3=0,\\
&&\iota k c_3=0,\\
&&c_1\left(\frac{-\iota \omega}{\alpha}+\iota k
u\right)+c_2\left\{(1+\gamma^2 u^2)\iota
k-(1-2\gamma^2u^2)(1+\gamma^2u^2)\frac{u'}{u}
-\frac{i\omega}{\alpha}\gamma^2 u\right\}=0,\\
&&c_1\gamma^2 \{a_z+uu'(1+\gamma^2 u^2)\}
+c_2[\gamma^2(1+\gamma^2u^2)\left(\frac{-\iota \omega}{\alpha}
+\iota u k\right)+\nonumber\\
&&+\gamma^2\{u'(1+\gamma^2 u^2)(1+4\gamma^2 u^2) +2u\gamma^2
a_z\}]=0.
\end{eqnarray}
Eqs.(3.15) and (3.16) show that $c_3$ is zero which implies that
there are no perturbations occurring in magnetic field of the
fluid. Thus we are left with Eqs.(3.17) and (3.18). It is
mentioned here that for the non-magnetized plasma, we also obtain
the same two equations.

\subsection{Numerical Solutions}

We consider time lapse $\alpha=z$ and $\rho$ as a constant. Using
mass conservation law in three dimensions, we obtain the value of
$u=\frac{1}{\sqrt{1+z^2}}$. When we use these values, the
determinant of the coefficients of constants $c_1$ and $c_2$ in
Eqs.(3.17) and (3.18) give us a complex number. Equating it to
zero, we obtain three values for $k$, two from the real part and
one from the imaginary part. Using the value of $k$, one can
calculate the following quantities:
\begin{enumerate}
  \item \textbf{Phase Velocity ($v_p$)}: The velocity with which
  the carrier wave in a modulated signal moves and can be
  calculated by $\frac{\omega}{k}$.
  \item  \textbf{Refractive Index ($n$)}: A property of a material that
  changes the speed of light, computed as the ratio of the speed of light
  in a vacuum to the speed of light through the material.
  Refractive index is the inverse of the phase velocity.
  \item \textbf{Change in refractive index with respect to angular frequency
  ($\frac{dn}{d\omega}$)}:
  This term helps to find out whether the dispersion is normal
  or not.
  \item \textbf{Group Velocity ($v_g$)}: This is the speed of transmission
  of information and/or
  energy in a wave packet and can be calculated by the formula
  $\frac{1}{n+\omega\frac{dn}{d\omega}}$.
\end{enumerate}
We would like to mention here that only longitudinal oscillations
are calculated in this case as we are using the longitudinal part
of the momentum equation. Thus we would discuss the longitudinal
waves propagating parallel to the magnetic field $\textbf{B}$ in
the remaining part of this section.

The dispersion equation from real part of the matrix is of the
form $A(z)k^2+B(z)k\omega+C(z)\omega^2+D(z)=0.$ This leads to two
values of $k$ in terms of $z$ and angular frequency shown in the
Figures 1 and 2. These are conjugate roots of quadratic equation.
The real wave numbers are shown in the graph lying in the region
$3.5\times 10^{-7}\leq z\leq 1.41.$
\begin{figure}
\center \epsfig{file=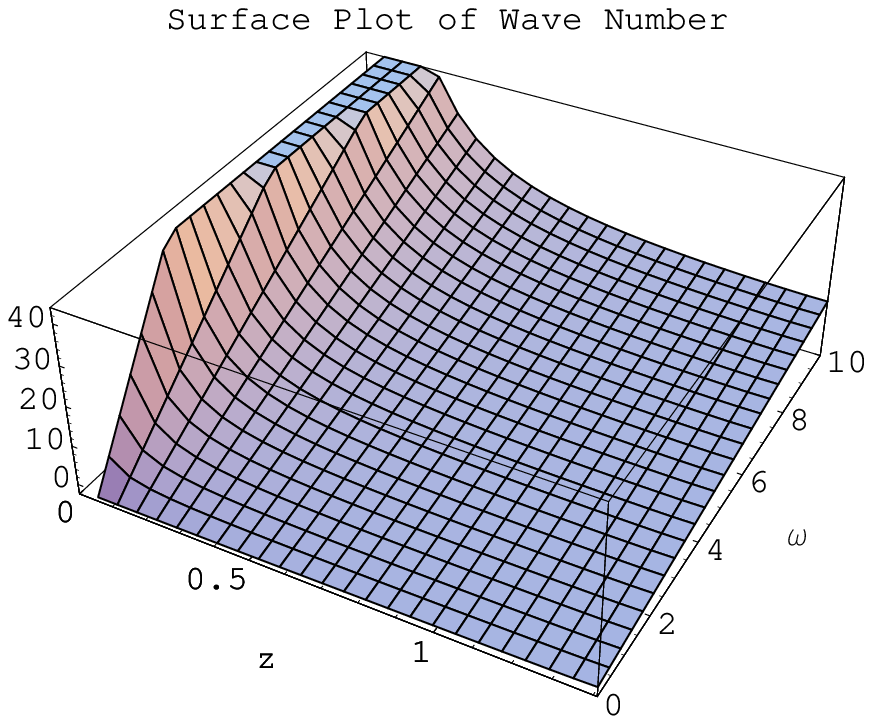,width=0.4\linewidth} \center
\epsfig{file=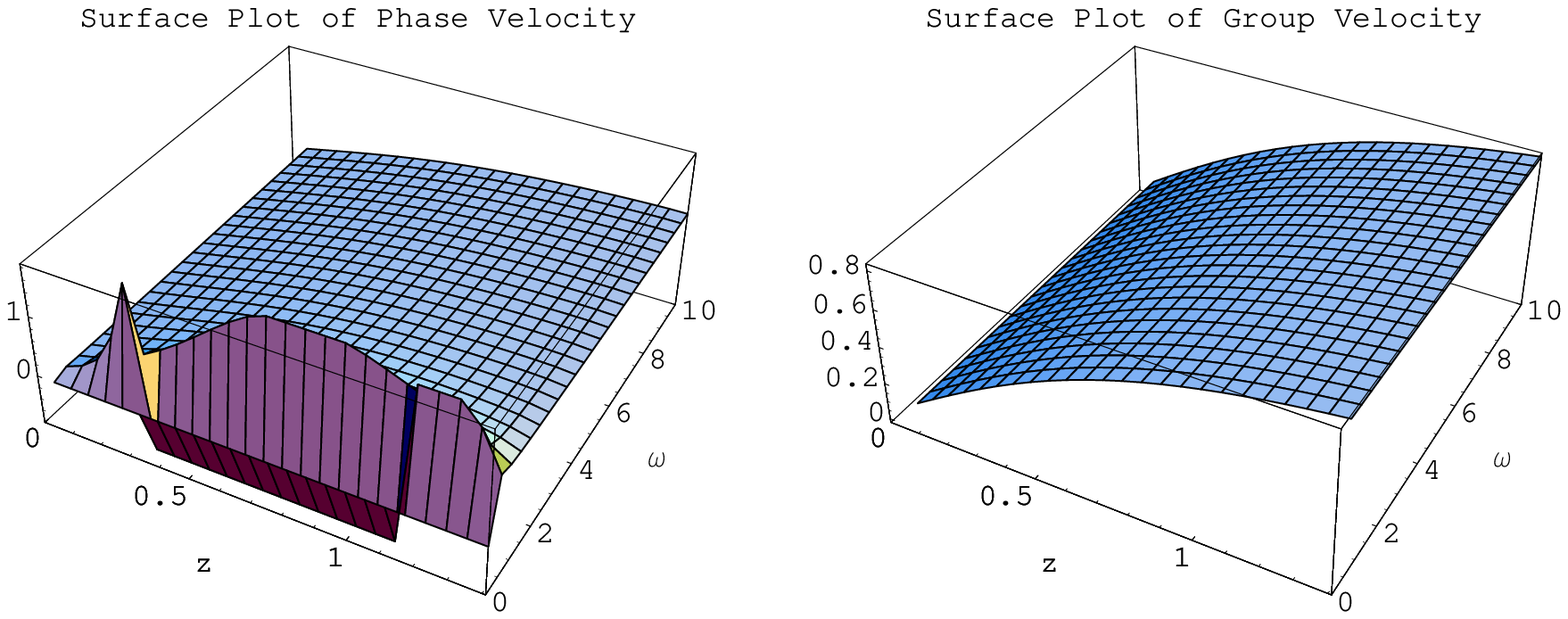,width=0.8\linewidth} \center
\epsfig{file=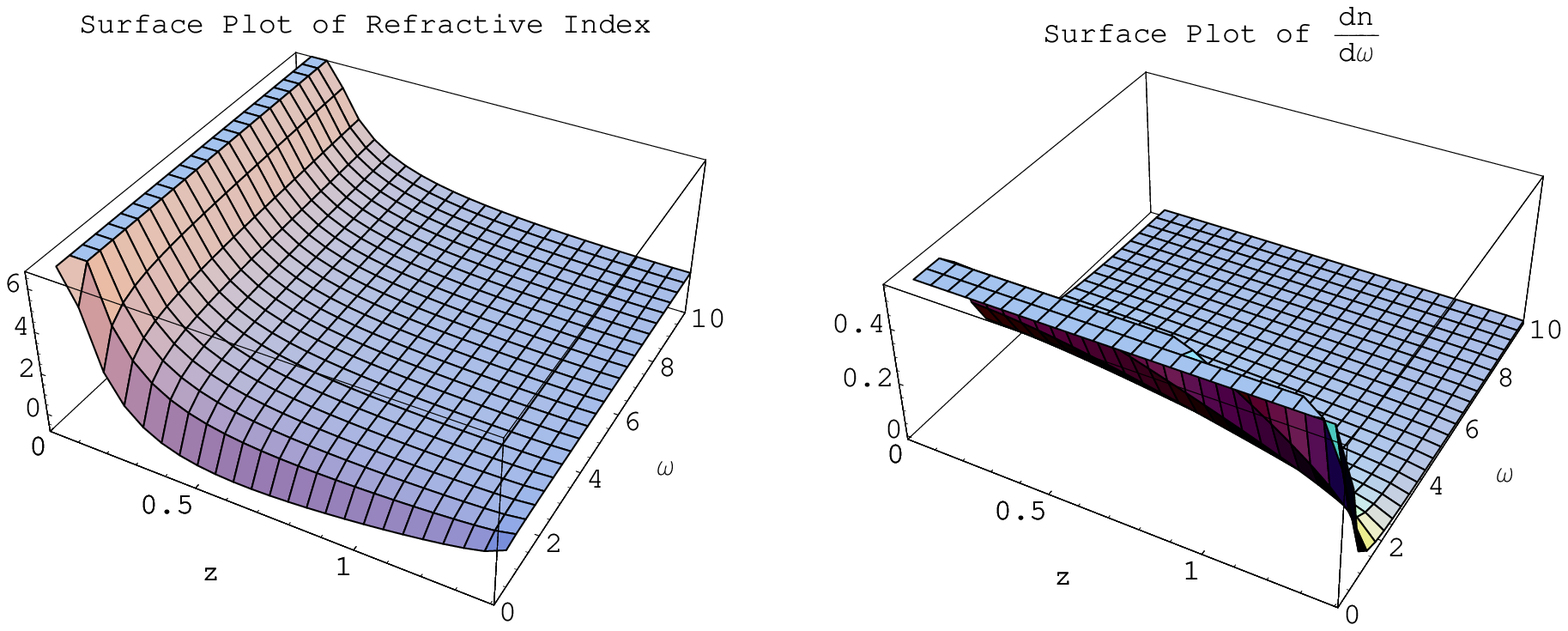,width=0.8\linewidth} \caption{The waves
decrease as they move away from the event horizon. The phase and the
group velocities are equal except for the waves with less angular
frequencies. The region shows normal dispersion.}
\end{figure}

We see from the Figure 1 that angular frequency $\omega$ and wave
number $k$ are directly proportional to each other, i.e., the
change in angular velocity gives the respective change in wave
number. Also, the wave number decreases as $z$ increases. This
means that waves lose energy as we go away from the black hole
horizon. At the horizon, the wave number becomes very large and
the waves vanish due to the effect of gravity. There is an abrupt
increment in phase velocity when the angular frequency is near to
zero. The region $3.5\times 10^{-7}\leq
z\leq0.5,~1.5\leq\omega\leq10$ has refractive index greater than
one and its variation with respect to omega greater than zero.
This corresponds to the region of normal dispersion [30]. The
group and wave velocities go parallel and are increasing as we go
away from the horizon.
\begin{figure}
\center \epsfig{file=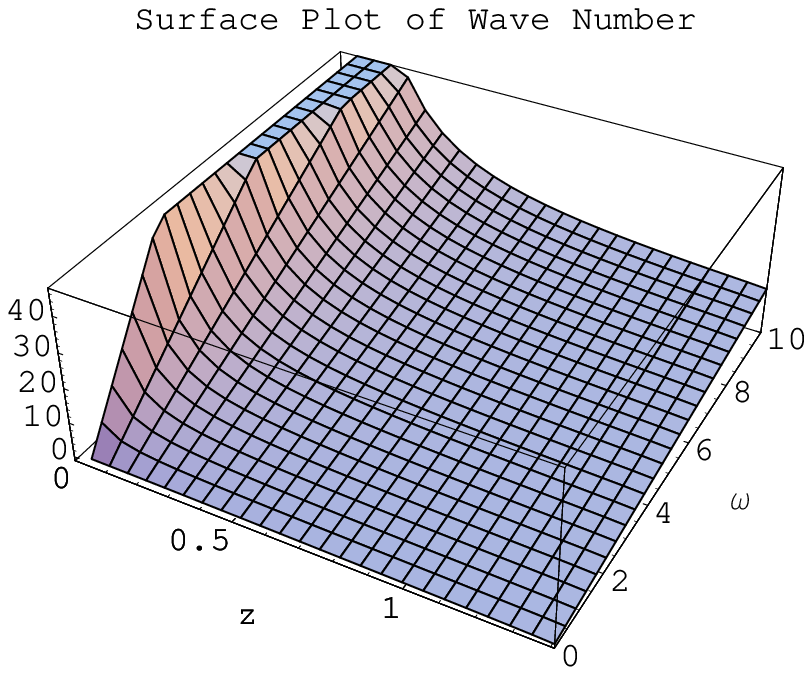,width=0.4\linewidth} \center
\epsfig{file=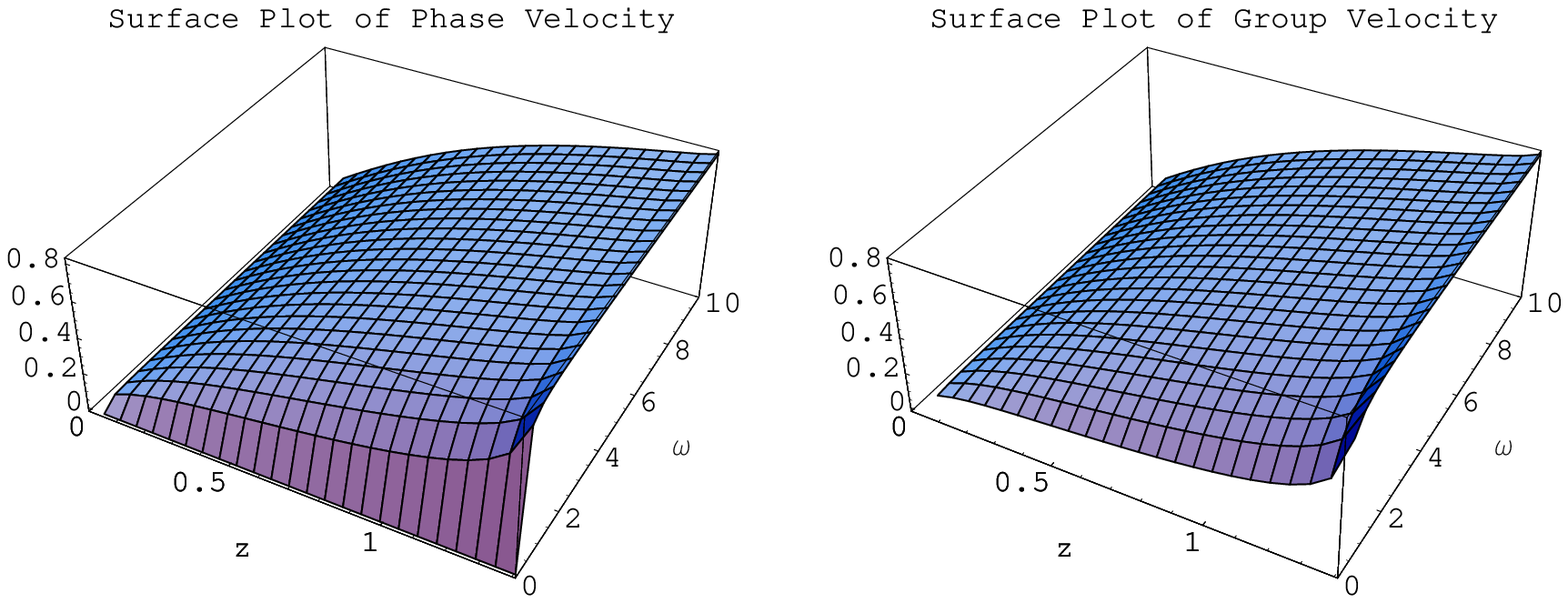,width=0.8\linewidth} \center
\epsfig{file=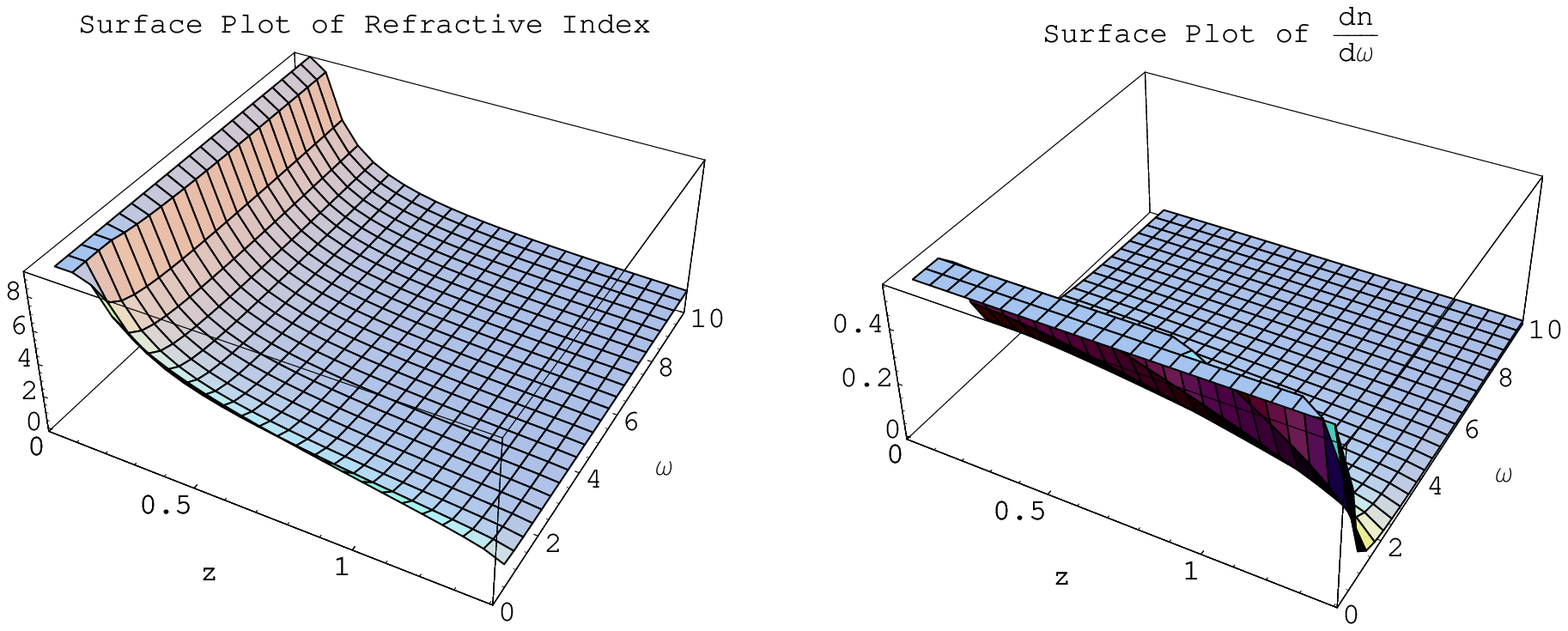,width=0.8\linewidth} \caption{The waves damp as
they go away from the event horizon. The phase and the group
velocities are equal except for the waves with negligible angular
frequencies. The region shows normal dispersion.}
\end{figure}

The Figure 2 shows that the waves are growing with the increase in
angular velocity, but damping occurs with as $z$ increases. For
$0\leq z<3.5\times10^{-7}$, the wave number goes to infinity and
hence there is no wave at the horizon. This shows that the event
horizon is the interface at which they are formed and as we go
away from the event horizon, the intensity of these waves decay.
Since the refractive index is greater than one and
$\frac{dn}{d\omega}$ is greater than zero, the whole region shows
normal dispersion.

The dispersion equation obtained from the imaginary part of the
matrix is of the form $A(z)k+B(z)\omega=0.$ Hence we are with only
one value of $k$ given by the Figure 3.
\begin{figure}
\center \epsfig{file=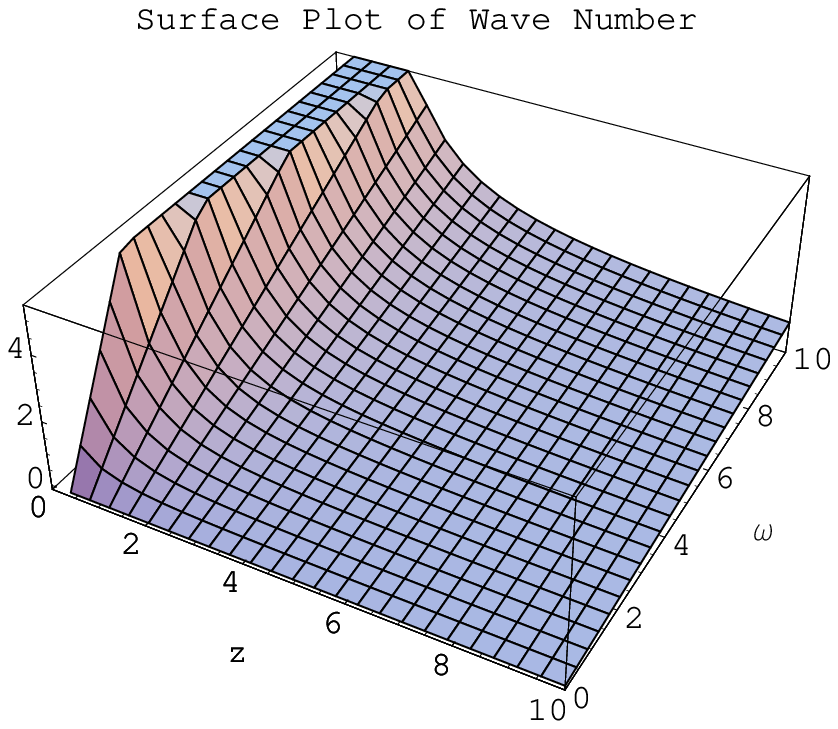,width=0.4\linewidth} \center
\epsfig{file=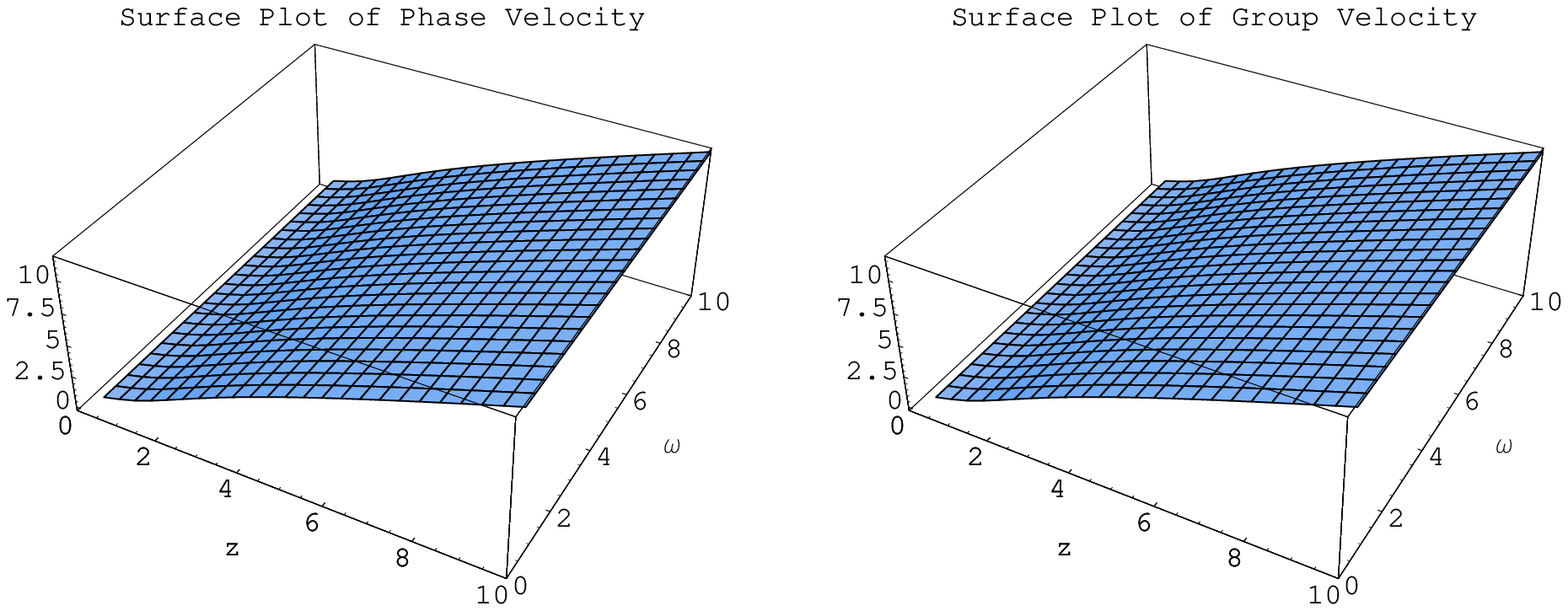,width=0.8\linewidth} \center
\epsfig{file=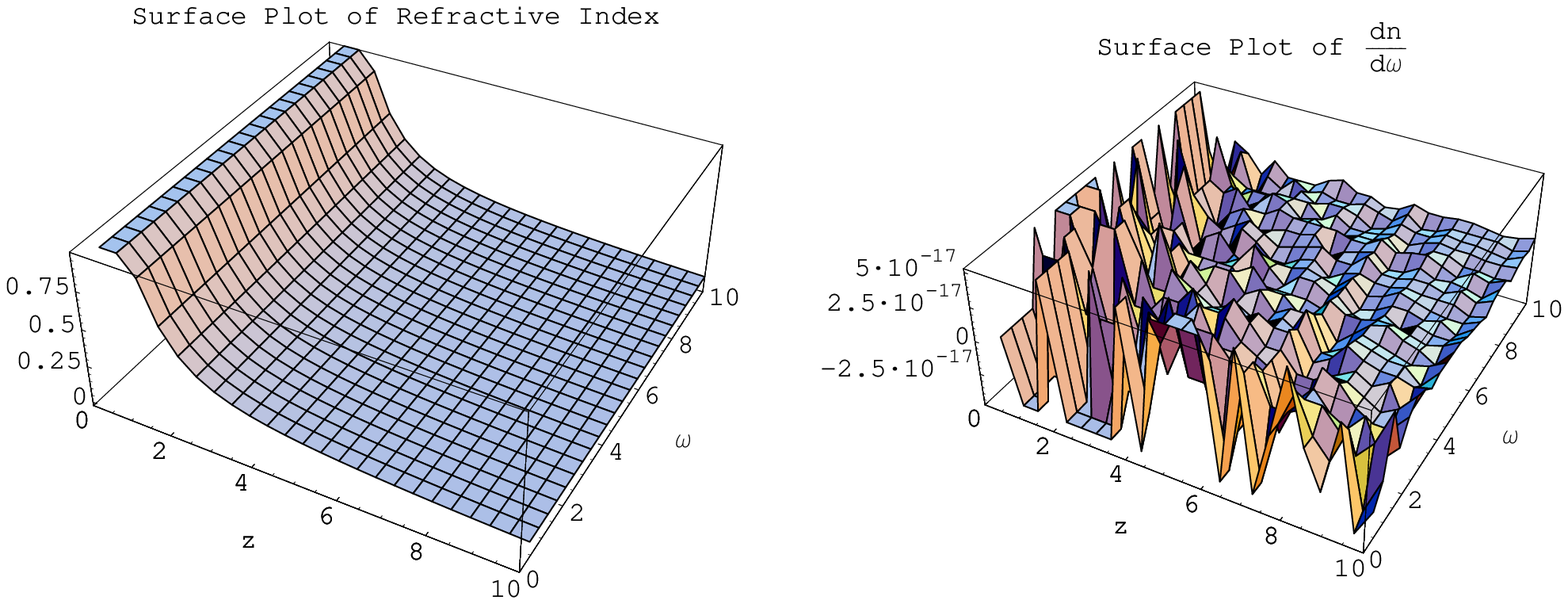,width=0.8\linewidth} \caption{The waves damp as
they move away from the event horizon. The phase and the group
velocities are same. The dispersion is not normal.}
\end{figure}

The Figure 3 shows that the waves gain energy with the increase in
angular velocity, but lose as $z$ increases. The wave number goes
to infinity and hence there is no wave for $0\leq
z<3.5\times10^{-7}$. The group and phase velocities are showing
the same behavior i.e. they are increasing with an increase in
$z$. The refractive index is not greater than one and hence the
dispersion is not normal.

\section{Rotating Non-Magnetized Background}

When we consider non-magnetized cold plasma in rotating
background, i.e., $\textbf{B}=0$, the GRMHD Eqs.(2.17) and (2.18)
vanish. Eqs.(2.19) and (2.20) change into general relativistic
hydrodynamical equations given below:
\begin{eqnarray}{\setcounter{equation}{1}}
&&\frac{\partial \rho}{\partial t}+(\alpha \textbf{V}.\nabla)\rho
+\rho\gamma^2\textbf{V}.\frac{\partial \textbf{V}}{\partial
t}+\rho\gamma^2\textbf{V}.(\alpha \textbf{V}.\nabla)\textbf{V}+
\rho \nabla.(\alpha\textbf{V})=0,\\
&&(\rho \gamma^2\delta_{ij}+\rho
\gamma^4V_iV_j)\left(\frac{1}{\alpha}\frac{\partial}{\partial
t}+\textbf{V}.\nabla\right)V^j=-\rho \gamma^2 a_i.
\end{eqnarray}
For rotating background, the fluid four-velocity measured by FIDO
can be described in $xz$-plane as follows
\begin{equation}
\textbf{V}=V(z)\textbf{e}_\textbf{x}+u(z)\textbf{e}_\textbf{z}
\end{equation}
and the Lorentz factor $\gamma$ takes the form
\begin{equation}
\gamma=\frac{1}{\sqrt{1-u^2-V^2}}.
\end{equation}
Here we shall use the following notations for the perturbed
quantities
\begin{eqnarray}
\tilde{\rho}&\equiv&\frac{\delta \rho}{\rho}=\tilde{\rho}(t,z)\nonumber\\
\textbf{v}&\equiv& \delta \textbf{V}=\textbf{v}
_\textbf{x}(t,z)\textbf{e}_\textbf{x}+\textbf{v}
_\textbf{z}(t,z)\textbf{e}_\textbf{z}
\end{eqnarray}
and the perturbations have harmonic space and time dependence,
i.e.,
\begin{equation}\tilde{\rho},~
\textbf{v}\sim e^{-i (\omega t-k z)}.
\end{equation}
The perturbed variables take the form
\begin{eqnarray}
\tilde{\rho}(t,z)=c_1e^{-\iota (\omega t-kz)},\nonumber\\
\textbf{v}_\textbf{z}(t,z)=c_2e^{-\iota (\omega t-kz)},\nonumber\\
\textbf{v}_\textbf{x}(t,z)=c_3e^{-\iota (\omega t-kz)}.
\end{eqnarray}

\subsection{Perturbation Equations}

Introducing linear perturbations in Eqs.(4.1) and (4.2), it
follows that
\begin{eqnarray}
&&\left(\frac{1}{\alpha}\frac{\partial}{\partial
t}+\textbf{V}.\nabla \right)\delta \rho + \rho \gamma^2
\textbf{V}.\left(\frac{1}{\alpha} \frac{\partial}{\partial t}+
\textbf{V}.\nabla\right)\textbf{v} -
\frac{\delta \rho}{\rho}(\textbf{V}.\nabla)\rho +\rho (\nabla.\textbf{v})\nonumber\\
&&=-2\rho \gamma^2 (\textbf{V}.\textbf{v})( \textbf{V}. \nabla) ln
\gamma -\rho \gamma^2 (\textbf{V}.\nabla \textbf{V}).\textbf{v}
+\rho (\textbf{v}.\nabla ln u),\\
&&(\rho\gamma^2 \delta_{ij}+\rho\gamma^4 V_i V_j)\frac{1}{\alpha}
\frac{\partial v^j}{\partial t}+\rho \gamma^2v_{i,j}V^j +\rho
\gamma^4 V_i v_{j,k} V^j V^k=-\gamma^2\{\delta \rho+2 \rho
\gamma^2 (\textbf{V}.\textbf{v})\}a_i\nonumber\\&&
-\rho\gamma^4(v_i V^j + v^j V_i)V_{k,j}V^k- \gamma^2\{\delta
\rho V^j +2 \rho \gamma^2 (\textbf{V}.\textbf{v}) V^j +\rho v^j\} V_{i,j}\nonumber\\
&&-\gamma^4 V_i\{\delta \rho V^j+4 \rho \gamma^2
(\textbf{V}.\textbf{v})V^j+\rho v^j\}V_{j,k}V^k.
\end{eqnarray}
The component form of Eqs.(4.8) and (4.9) gives the following
three equations
\begin{eqnarray}
&&\frac{1}{\alpha}\frac{\partial \tilde{\rho}}{\partial t}+u
\tilde{\rho}_{,z}+\gamma^2 V\left(\frac{1}{\alpha} \frac{\partial
v_x}{\partial t}+uv_{x,z}\right)+\gamma^2 u\frac{1}{\alpha}
\frac{\partial v_z}{\partial t}+(1+\gamma^2 u^2)v_{z,z}\nonumber\\
&&=-\gamma^2 u\{(1+2\gamma^2 V^2)V'+2\gamma^2uVu'\}v_x
+\left\{(1-2\gamma^2u^2)(1+\gamma^2 u^2)\frac{u'}{u}-2\gamma^4u^2VV'\right\}v_z,\\
&& \gamma^2(1+\gamma^2V^2)\left(\frac{1}{\alpha} \frac{\partial
v_x}{\partial t}+u v_{x,z}\right)+\gamma^4uV(\frac{1}{\alpha}
\frac{\partial v_z}{\partial t}+uv_{z,z})\nonumber\\
&&=-\tilde{\rho}\gamma^2u\{(1+\gamma^2V^2)V'+\gamma^2uVu'\}
-\gamma^4u\{(1+4\gamma^2V^2)uu'+4VV'(1+\gamma^2V^2)\}v_x\nonumber\\
&&-[\gamma^2\{(1+2\gamma^2u^2)(1+2\gamma^2V^2)-\gamma^2V^2\}V'
+2\gamma^4(1+2\gamma^2u^2)uVu']v_z,\\
&&\gamma^2(1+\gamma^2 u^2)\left(\frac{1}{\alpha} \frac{\partial
v_z}{\partial t}+u v_{z,z}\right)+\gamma^4uV\left(\frac{1}{\alpha}
\frac{\partial v_x}{\partial t}+u v_{x,z}\right)\nonumber\\
&&=-\tilde{\rho}\gamma^2\{a_z+(1+\gamma^2u^2)uu'+\gamma^2 u^2 VV'\}\nonumber\\
&&-[\gamma^4\{u^2V'(1+4\gamma^2V^2)+2V(a_z+(1+2\gamma^2u^2)uu')\}]v_x\nonumber\\
&&-\gamma^2[u'(1+\gamma^2 u^2)(1+4\gamma^2 u^2)
+2u\gamma^2\{a_z+(1+2\gamma^2u^2)VV'\}]v_z.
\end{eqnarray}
Using Eq.(4.7), the Fourier analyzed form of Eqs.(4.10)-(4.12)
become
\begin{eqnarray}
&&c_1\left(-\frac{\iota \omega}{\alpha}+\iota k
u\right)+c_2\left[-\frac{\iota
\omega}{\alpha}\gamma^2u+(1+\gamma^2 u^2)\iota
k-(1-2\gamma^2u^2)(1+\gamma^2u^2)\frac{u'}{u}+
2\gamma^4u^2VV'\right]\nonumber\\
&&+c_3\gamma^2\left[-\frac{\iota \omega}{\alpha}V+\iota
kuV+\gamma^2u\{(1+2\gamma^2V^2)V'+2\gamma^2uVu'\}\right]=0,\\
&&c_1\gamma^2u\{(1+\gamma^2V^2)V'+\gamma^2uVu'\}+c_2\gamma^2[\left(-\frac{\iota
\omega}{\alpha}+\iota ku \right)\gamma^2uV\nonumber\\
&&+\{(1+2\gamma^2u^2)(1+2\gamma^2V^2)-\gamma^2V^2\}V'
+2\gamma^2(1+2\gamma^2u^2)uVu']\nonumber\\
&&+c_3[\left(-\frac{\iota \omega}{\alpha}+\iota ku\right)
\gamma^2(1+\gamma^2V^2)+\gamma^4u\{(1+4\gamma^2V^2)uu'+4VV'(1+\gamma^2V^2)\}]=0,\\
&&c_1\gamma^2\{a_z+(1+\gamma^2u^2)uu'+\gamma^2u^2VV'\}
+c_2[\gamma^2(1+\gamma^2u^2)\left(-\frac{\iota
\omega}{\alpha}+\iota uk\right)\nonumber\\
&&+\gamma^2\{u'(1+\gamma^2u^2)(1+4\gamma^2u^2)
+2u\gamma^2(a_z+(1+2\gamma^2u^2)VV')\}]\nonumber\\
&&+c_3\gamma^4\left[\left(-\frac{\iota \omega}{\alpha}+\iota u
k\right)uV+u^2V(1+4\gamma^2V^2)+2V\{a_z+uu'(1+2\gamma^2u^2)\}\right]=0.
\end{eqnarray}
The determinant of the coefficients of $c_1,~c_2$ and $c_3$ gives
a complex number. Equating it to zero yields a complex dispersion
relation in $k$.

\subsection{Numerical Solutions}

We consider time lapse $\alpha=z$ and assume that $V=u$ for the
sake of convenience. Using mass conservation law in three
dimensions (with mass density as a constant quantity) we obtain
$u=\frac{1}{\sqrt{2+z^2}}.$ In this section we are having modes
when $\textbf{B}=0$. The complex dispersion relation obtained from
the determinant of coefficients leads to two dispersion equations.
The real part gives an equation of the type
$A(z)k^2+B(z)k\omega+C(z)\omega^2+D(z)=0$ leading to two
dispersion relations shown in the Figures 4 and 5. The imaginary
part shows a cubic equation
$A(z)k^3+B(z)k^2\omega+C(z)k\omega^2+D(z)\omega^3=0.$ This
equation has one real and two complex conjugate solutions. The
Figure 6 shows the real value of $k$ obtained from this equation.
\begin{figure}
\center \epsfig{file=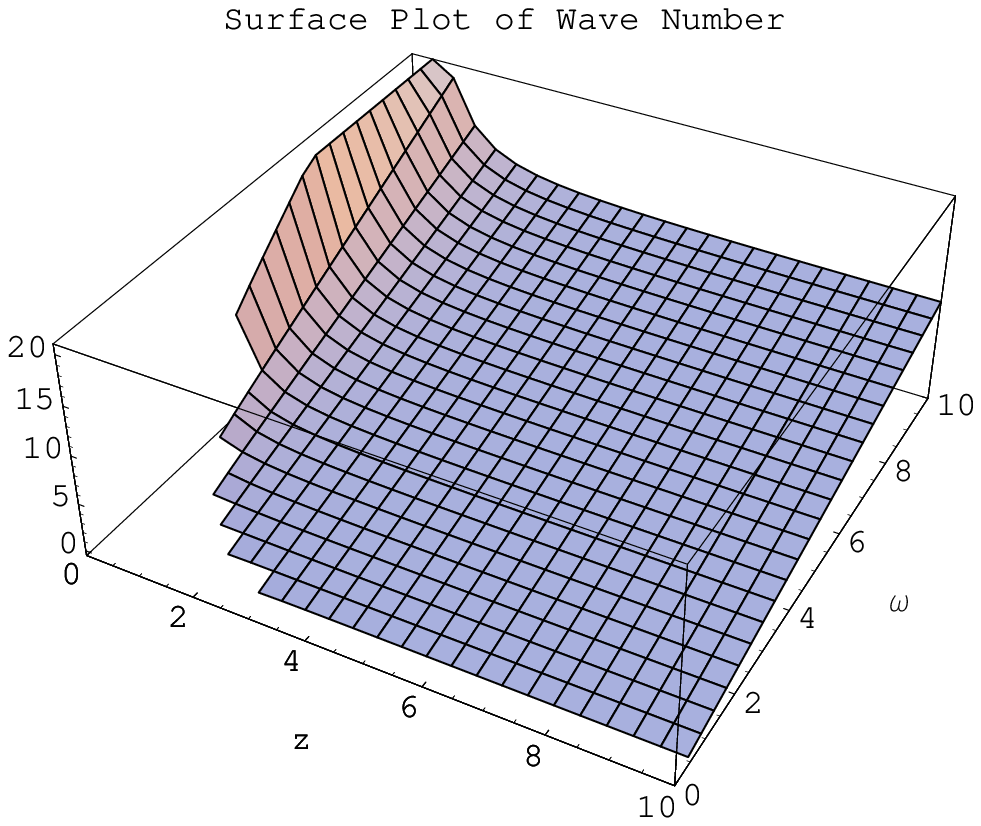,width=0.4\linewidth} \center
\epsfig{file=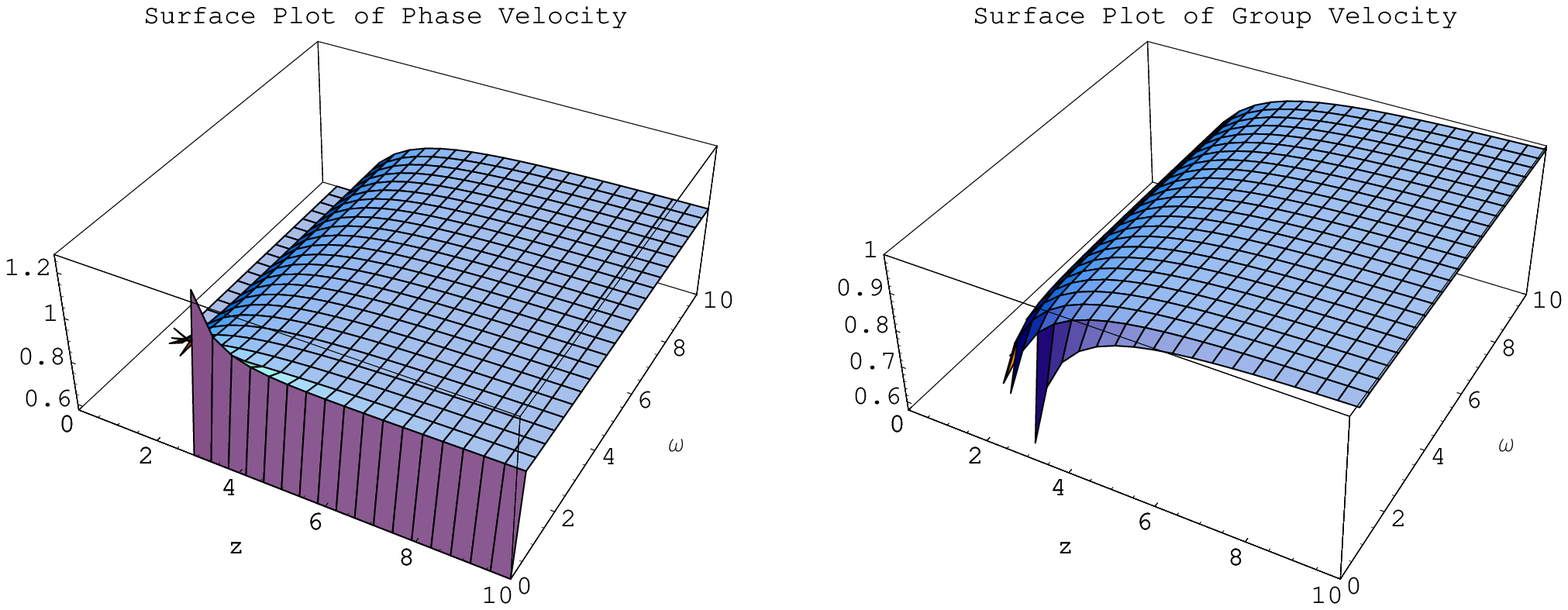,width=0.8\linewidth} \center
\epsfig{file=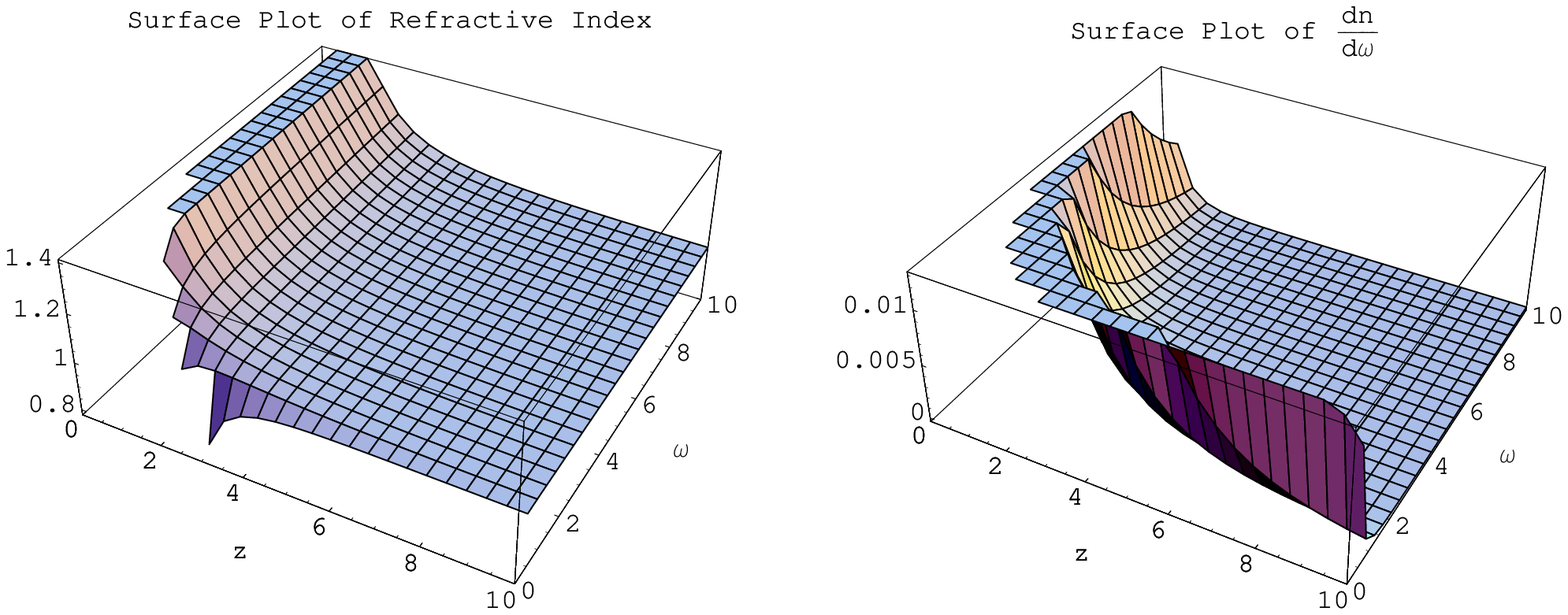,width=0.8\linewidth} \caption{The waves damp
when they move away from the event horizon. The phase velocity is
less than the group velocity in most of the region. Most of the
region shows normal dispersion.}
\end{figure}

The Figure 4 indicates that the region for which real waves exist
lies in $1.5\times 10^{-8}<z\leq10,~0.045<\omega\leq10$. It is
clear from the Figure that the region $0<\omega\leq 0.045$
contains evanescent waves. We note that at the horizon and nearby
values of $z$ i.e. in the region $0\leq z\leq 1.5\times 10^{-8}$,
the wave number is infinite and hence no wave exists there. The
wave number is decreasing as we go away from the horizon and hence
damping occurs. Also $n>1,~\frac{dn}{d\omega}>0$ and hence the
region is of normal dispersion. The phase velocity is less than
group velocity of waves except at some points where the refractive
index $n$ is positive but is less than one, as shown in the graph.
This means that the dispersion is not normal but anomalous at
these points [31], normal otherwise. At $\omega=0$ and very small
angular frequencies, the phase velocity is high but the group
velocity become infinite which is not physical.
\begin{figure}
\center \epsfig{file=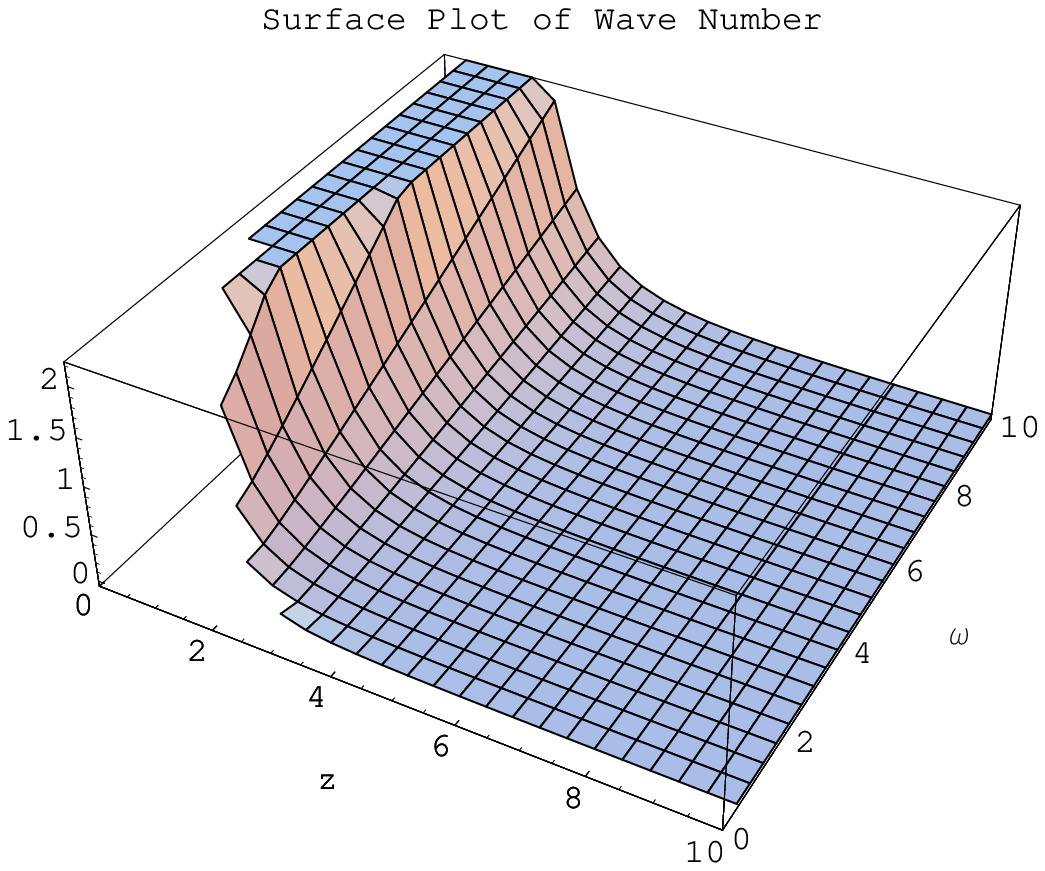,width=0.4\linewidth} \center
\epsfig{file=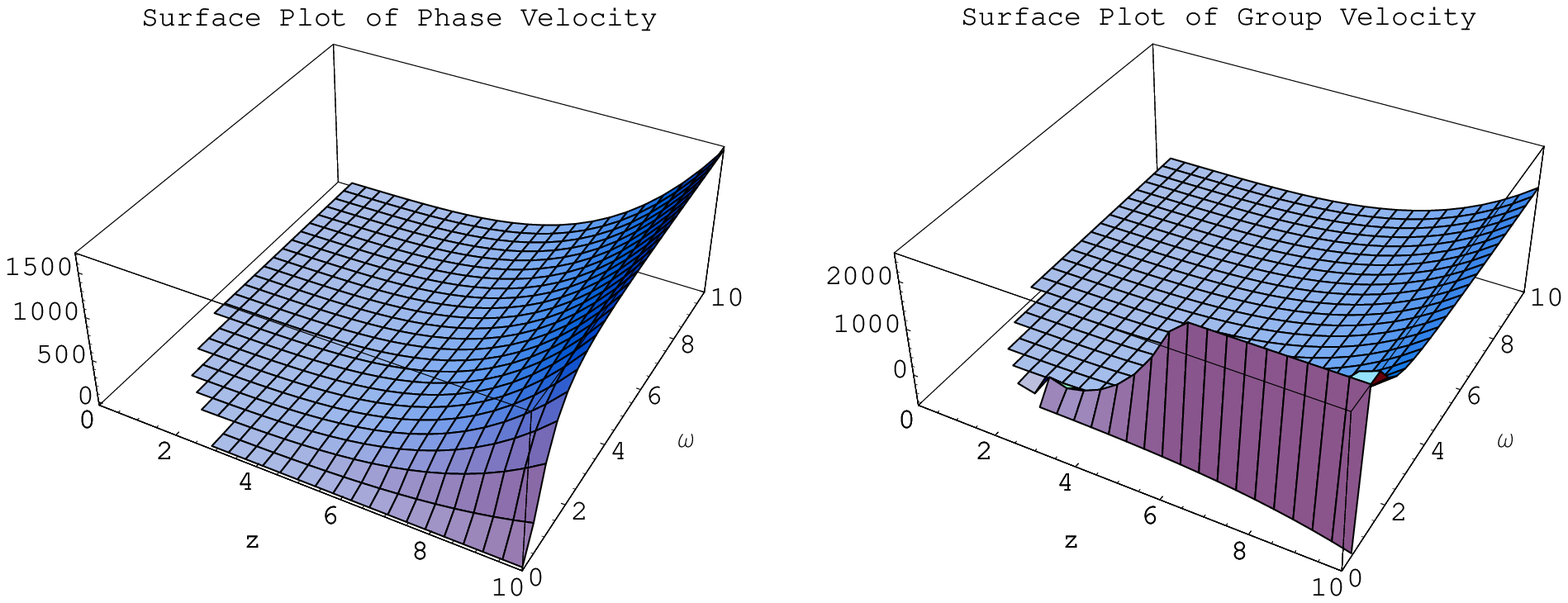,width=0.8\linewidth} \center
\epsfig{file=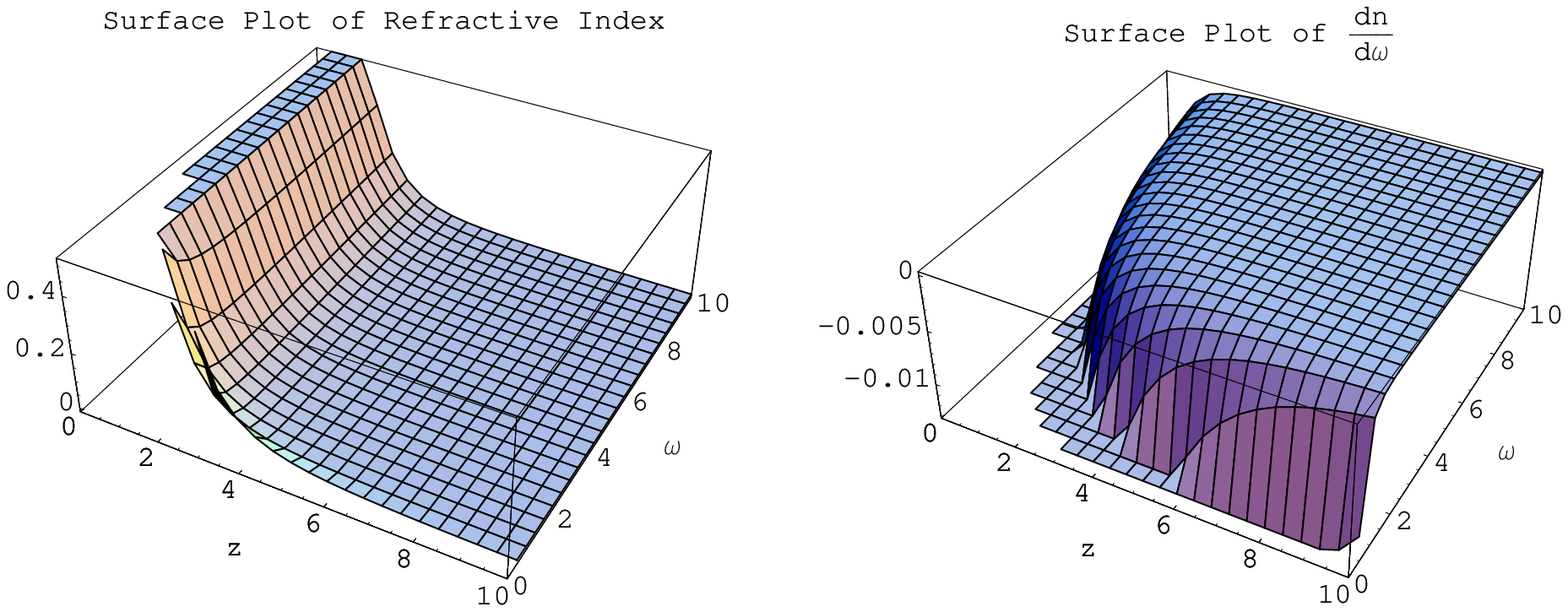,width=0.8\linewidth} \caption{The waves damp
while they move away from the event horizon. The dispersion is not
normal.}
\end{figure}

The Figure 5 contains the infinite wave number in the region
$0\leq z<1.5\times 10^{-8}$, that is no wave present there. The
wave number possesses real values in the region $1.5\times
10^{-8}<z\leq10,~0.05\leq\omega\leq10$. On increasing the value of
$z$, the waves are damping but are growing with the increase of
the angular frequency. In this region, refractive index $n<1$ and
$\frac{dn}{d\omega}<0$ which implies that the region is not of
normal dispersion. At $1.5\times
10^{-8}<z\leq10,~0\leq\omega\leq4$, the wave number becomes
imaginary at several points leading to the result that evanescent
waves present there.
\begin{figure} \center
\epsfig{file=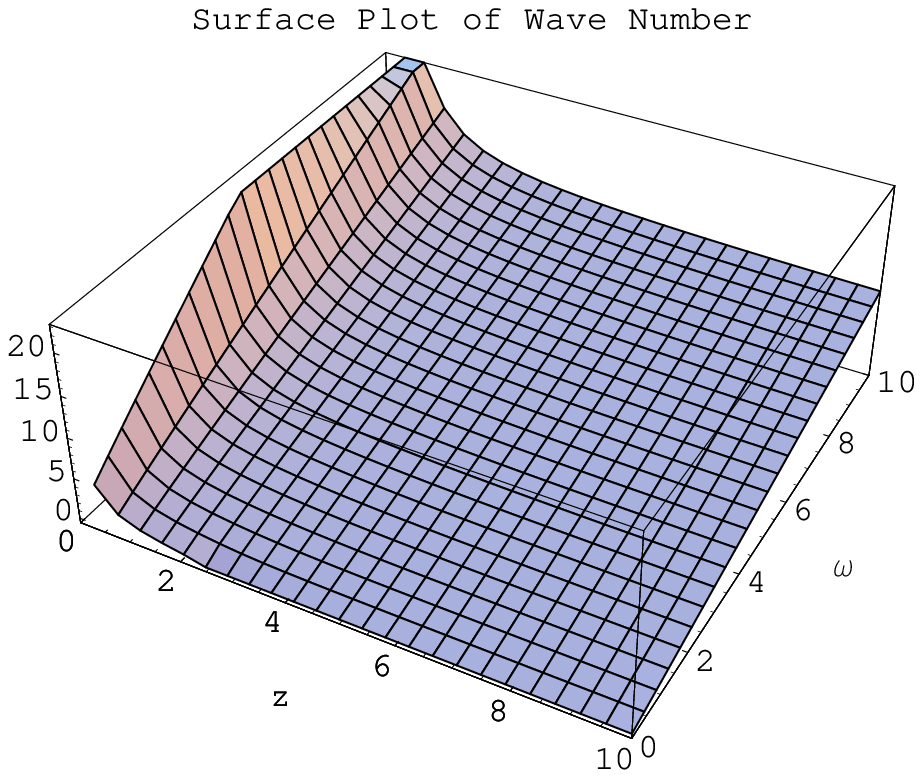,width=0.4\linewidth} \center
\epsfig{file=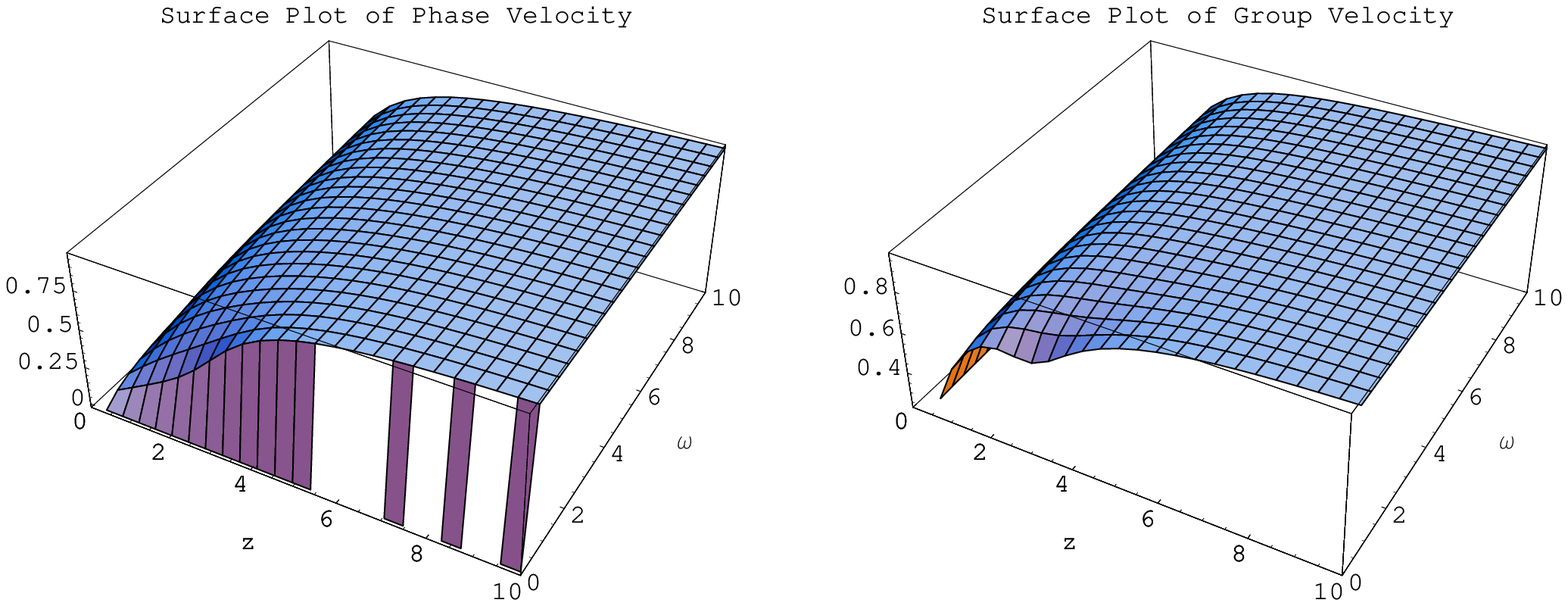,width=0.8\linewidth} \center
\epsfig{file=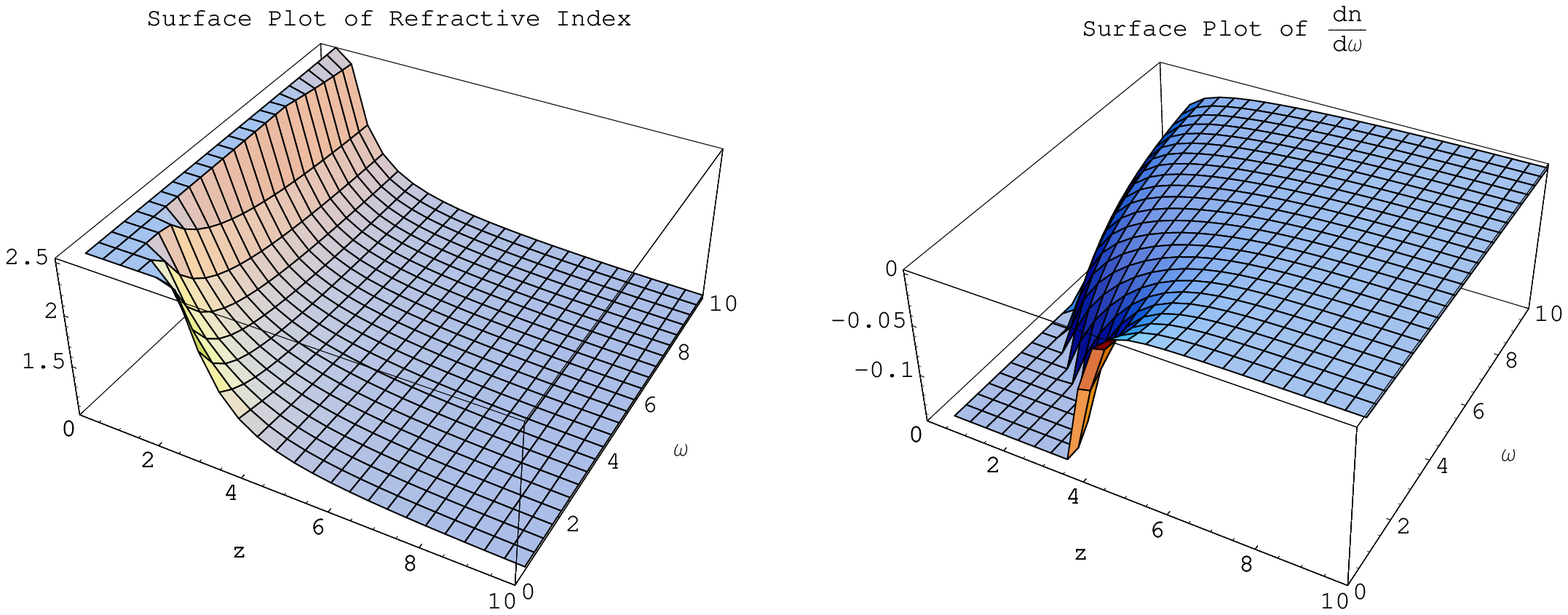,width=0.8\linewidth} \caption{The waves damp as
they move away from the event horizon. The group velocity is greater
than the phase velocity. The region is of anomalous dispersion.}
\end{figure}

In the Figure 6, real waves exist in the region $1.5\times
10^{-8}\leq z\leq 10.$ The refractive index is greater than one,
$\frac{dn}{d\omega}$ is negative and thus the region is not of
normal dispersion. No wave is present on the event horizon because
the wave number is infinite there. As we go away from the horizon,
the wave number decreases. On increasing the angular frequency, we
find growing waves. Also, group velocity is greater than phase
velocity which indicates that the region is of anomalous
dispersion.

\section{Rotating Magnetized Background}

The GRMHD equations for magnetized cold plasma in rotating
background will remain the same as given by Eqs.(2.17)-(2.20).
Here the four-velocity of the fluid measured by FIDO can be
described in the same way as given in the previous section. The
rotating magnetic field can be expressed in the $xz$-plane as
\begin{equation}{\setcounter{equation}{1}}
\textbf{B}=B[\lambda(z)\textbf{e}_\textbf{x}+\textbf{e}_\textbf{z}],
\end{equation}
where $\lambda,~u$ and $V$ are related to each other by
\begin{equation}
V=\frac{V_F}{\alpha}+\lambda u,
\end{equation}
$V_F$ as an integration constant. The following notations for the
perturbed magnetic field will be used in addition to the notations
given by Eq.(4.5).
\begin{eqnarray}
\textbf{b}&\equiv& \frac{\delta \textbf{B}}{B}=\textbf{b}
_\textbf{x}(t,z)\textbf{e}_\textbf{x}+\textbf{b}
_\textbf{z}(t,z)\textbf{e}_\textbf{z}.
\end{eqnarray}
Since the perturbations have harmonic space and time dependence,
i.e., $\textbf{b}\sim e^{-i (\omega t-k z)}$, the perturbed
variable can be expressed as follows
\begin{eqnarray}
\textbf{b}_\textbf{x}(t,z)=c_4e^{-\iota (\omega t-kz)},\nonumber\\
\textbf{b}_\textbf{z}(t,z)=c_5e^{-\iota (\omega t-kz)}.
\end{eqnarray}

\subsection{Perturbation Equations}

Using Eq.(2.21) in Eqs.(2.17)-(2.20), we have
\begin{eqnarray}
&&\frac{\partial (\delta \textbf{B})}{\partial t}=\nabla
\times(\alpha \textbf{v} \times \textbf{B})
+\nabla \times (\alpha \textbf{V} \times \delta \textbf{B}),\\
&&\nabla. (\delta \textbf{B})=0,\\
&&\left(\frac{1}{\alpha}\frac{\partial}{\partial
t}+\textbf{V}.\nabla \right)\delta \rho + \rho \gamma^2
\textbf{V}.\left(\frac{1}{\alpha} \frac{\partial}{\partial t}+
\textbf{V}.\nabla\right)\textbf{v}
-\frac{\delta \rho}{\rho}(\textbf{V}.\nabla)\rho +\rho (\nabla.\textbf{v})\nonumber\\
&&=-2\rho \gamma^2 (\textbf{V}.\textbf{v})( \textbf{V}. \nabla) ln
\gamma -\rho \gamma^2 (\textbf{V}.\nabla \textbf{V}).\textbf{v}
+\rho(\textbf{v}.\nabla ln u),\\
&&\left\{\left(\rho\gamma^2+\frac{\textbf{B}^2}{4\pi}\right)\delta_{ij}+\rho\gamma^4
V_i V_j-\frac{1}{4 \pi}B_i B_j\right\}\frac{1}{\alpha}
\frac{\partial v^j}{\partial t}+\frac{1}{4 \pi}\left[\textbf{B}
\times \left\{\textbf{V} \times
\frac{1}{\alpha}\frac{\partial(\delta \textbf{B})}
{\partial t}\right\}\right]_i\nonumber\\
&&+\rho \gamma^2v_{i,j}V^j +\rho \gamma^4 V_i v_{j,k} V^j V^k
-\frac{1}{4 \pi \alpha}\{(\alpha \delta B_i)_{,j}-(\alpha \delta
B_j)_{,i}\} B^j=-\gamma^2\{\delta \rho+2 \rho \gamma^2
(\textbf{V}.\textbf{v})\}a_i\nonumber\\
&&+\frac{1}{4 \pi \alpha}\{(\alpha B_i)_{,j}-(\alpha
B_j)_{,i}\}\delta B^j-\rho\gamma^4 (v_i V^j + v^j V_i)V_{k,j}V^k-
\gamma^2\{\delta
\rho V^j +2 \rho \gamma^2 (\textbf{V}.\textbf{v}) V^j +\rho v^j\} V_{i,j}\nonumber\\
&&-\gamma^4 V_i\{\delta \rho V^j+4 \rho
\gamma^2(\textbf{V}.\textbf{v})V^j+\rho v^j\}V_{j,k}V^k.
\end{eqnarray}
The component form of the Eqs.(5.5)-(5.8) can be written, after a
tedious algebra, as follows
\begin{eqnarray}
&&\frac{1}{\alpha}\frac{\partial b_x}{\partial t}+ub_{x,z}=\nabla
ln \alpha(v_x-\lambda v_z +V b_z-u
b_x)+(v_{x,z}-\lambda v_{z,z}-\lambda' v_{z}+V'b_z-u'b_x),\\
&&\frac{1}{\alpha}\frac{\partial b_z}{\partial
t}+ub_{z,z}=0,\\
&&b_{z,z}=0,\\
&&\frac{1}{\alpha}\frac{{\partial \tilde{\rho}}}{\partial t}+u
\tilde{\rho}_{,z} +\gamma^2 V\frac{1}{\alpha}\frac{\partial
v_x}{\partial t}+\gamma^2 u\frac{1}{\alpha} \frac{\partial
v_z}{\partial t}+(1+\gamma^2 u^2)v_{z,z}+\gamma^2uVv_{x,z}\nonumber\\
&&=-\gamma^2u[(1+2\gamma^2 V^2)V'+2\gamma^2uVu']v_x
+[(1-2\gamma^2u^2)(1+\gamma^2u^2)\frac{u'}{u}-2\gamma^4u^2VV']v_z,\\
&&\left\{\rho
\gamma^2(1+\gamma^2V^2)+\frac{B^2}{4\pi}\right\}\frac{1}{\alpha}
\frac{\partial v_x}{\partial t}+\left(\rho \gamma^4 u
V-\frac{\lambda B^2}{4\pi}\right)\frac{1}{\alpha}\frac{\partial
v_z}{\partial t}
+\left\{\rho \gamma^2(1+\gamma^2 V^2)-\frac{B^2}{4\pi}\right\}u v_{x,z}\nonumber\\
&&+\left(\rho \gamma^4uV+\frac{\lambda
B^2}{4\pi}\right)uv_{z,z}-\frac{B^2}{4\pi}(1-u^2)b_{x,z}
-\frac{B^2}{4\pi\alpha}b_x \{\alpha'(1-u^2)-\alpha uu'\}\nonumber\\
&&+\tilde{\rho}\rho\gamma^2 u\{(1+\gamma^2
V^2)V'+\gamma^2uVu'\}+\left[\rho\gamma^4u\{(1+4\gamma^2 V^2)u
u'+4(1+\gamma^2
V^2)VV'\}+\frac{B^2 u \alpha'}{4\pi \alpha}\right]v_x\nonumber\\
&&+\left[\rho\gamma^2 [\{(1+2 \gamma^2 u^2)(1+2 \gamma^2 V^2) -
\gamma^2 V^2\}V'+2\gamma^2(1+2\gamma^2 u^2)u V u'] +\frac{B^2 u}{4
\pi\alpha}(\lambda\alpha)'\right]v_z=0,\\
&&\left\{\rho \gamma^2(1+\gamma^2 u^2)+ \frac{\lambda^2
B^2}{4\pi}\right\}\frac{1}{\alpha}\frac{\partial v_z}{\partial
t}+\left(\rho \gamma^4 u V-\frac{\lambda
B^2}{4\pi}\right)\frac{1}{\alpha}\frac{\partial v_x}{\partial t}
+\left\{\rho \gamma^2(1+\gamma^2 u^2)
-\frac{\lambda^2 B^2}{4\pi}\right\}u v_{z,z}\nonumber\\
&&+\left(\rho \gamma^4 u V+\frac{\lambda
B^2}{4\pi})uv_{x,z}\right)+\frac{\lambda B^2}{4
\pi}(1-u^2)b_{x,z}+\frac{B^2}{4\pi \alpha}\{(\alpha
\lambda)'+\alpha'\lambda-u\lambda(u\alpha'+u'\alpha)\}b_x\nonumber\\
&&+\tilde{\rho} \gamma^2[a_z+u\{(1+\gamma^2 u^2)u'+\gamma^2uVV'\}]\nonumber\\
&&+\left[\rho\gamma^4\{u^2V'(1+4\gamma^2V^2)+2V(a_z+uu'(1+2\gamma^2
u^2))\}+\frac{\lambda B^2 \alpha' u}{4\pi\alpha}\right]v_x\nonumber\\
&&+\left[\rho\gamma^2\{u'(1+\gamma^2 u^2)(1+4\gamma^2 u^2)
+2u\gamma^2\{(1+2\gamma^2 u^2)VV'+a_z\}\}-\frac{\lambda B^2
u}{4\pi\alpha}(\alpha\lambda)'\right]v_z=0.
\end{eqnarray}
When the above equations are Fourier analyzed, they take the
following form
\begin{eqnarray}
&&c_3(\alpha'+\iota k \alpha)-c_2\{(\alpha \lambda)' +\iota
k\alpha\lambda\}+c_5(\alpha V)'
-c_4\{(\alpha u)'-\iota \omega +\iota k \alpha u \}=0,\\
&&c_5\left(-\frac{\iota \omega}{\alpha}+\iota k u\right)=0,\\
&&c_5 \iota k=0,\\
&&c_1\left(-\frac{\iota \omega}{\alpha}+\iota k
u\right)+c_2\left\{-\frac{\iota \omega}{\alpha}\gamma^2 u+\iota
k(1+\gamma^2u^2)-(1-2\gamma^2u^2)(1+\gamma^2u^2)\frac{u'}{u}+2\gamma^4
u^2VV'\right\}\nonumber\\
&&+c_3\gamma^2\left[\left(-\frac{\iota\omega}{\alpha}+\iota
ku\right)V+u\{(1+2\gamma^2V^2)V'+2\gamma^2uVu'\}\right]=0,\\
&&c_1\rho\gamma^2u\{(1+\gamma^2V^2)V'+\gamma^2uVu'\}-\frac{B^2}{4\pi}c_4\{(1-u^2)\iota
k+\frac{\alpha'}{\alpha}(1-u^2)-uu'\}\nonumber\\
&&+c_2\left[-\left(\rho\gamma^4 u V- \frac{\lambda
B^2}{4\pi}\right)\frac{\iota \omega}{\alpha}+\iota k
u\left(\rho\gamma^4 u V+ \frac{\lambda
B^2}{4\pi}\right)+\rho\gamma^2\{(1+2\gamma^2 u^2)(1+2\gamma^2V^2)
-\gamma^2V^2\}V'\right.\nonumber\\
&&\left.+2\rho\gamma^4(1+2\gamma^2 u^2)u V
u'+\frac{B^2u}{4\pi\alpha}(\lambda \alpha)'\right]
+c_3\left[-\left\{\rho\gamma^2(1+\gamma^2 V^2)+
\frac{B^2}{4\pi}\right\}\frac{\iota \omega}{\alpha}\right.\nonumber\\
&&\left.+\iota ku\left\{\rho\gamma^2(1+\gamma^2
V^2)-\frac{B^2}{4\pi}\right\}+\rho\gamma^4u\{(1+4\gamma^2V^2)uu'+4(1+\gamma^2V^2)VV'\}
-\frac{B^2u\alpha'}{4\pi\alpha}\right]=0,\\
&&c_1\rho\gamma^2[a_z +u\{(1+\gamma^2 u^2)u'+\gamma^2VuV'\}]\nonumber\\
&&+c_2\left[-\left\{\rho \gamma^2(1+\gamma^2 u^2)+\frac{\lambda^2
B^2}{4\pi}\right\}\frac{\iota \omega}{\alpha}+\left\{\rho
\gamma^2(1+\gamma^2u^2)-\frac{\lambda^2 B^2}{4\pi}\right\}\iota
ku\right.\nonumber\\
&&\left.+\left\{\rho\gamma^2\{u'(1+\gamma^2 u^2)(1+4\gamma^2 u^2)
+2u\gamma^2 ((1+2\gamma^2 u^2)VV'+a_z)\}-\frac{\lambda B^2
u}{4\pi\alpha}(\alpha \lambda)'\right\}\right]\nonumber\\
&&+c_3\left[-\left(\rho \gamma^4 u V- \frac{\lambda
B^2}{4\pi}\right)\frac{\iota \omega}{\alpha}+\left(\rho \gamma^4 u
V+\frac{\lambda B^2}{4\pi}\right)\iota ku\right.\nonumber\\
&&\left.+\left\{\rho\gamma^4 \{u^2V'(1+4\gamma^2
V^2)+2V(a_z+uu'(1+2\gamma^2 u^2))\}+\frac{\lambda B^2 \alpha'
u}{4\pi\alpha}\right\}\right]\nonumber\\
&&+\frac{B^2}{4 \pi}c_4\left[\lambda (1-u^2)\iota
k+\lambda\frac{\alpha'}{\alpha}(1-u^2)-\lambda uu'+\frac{(\lambda
\alpha)'}{\alpha}\right]=0.
\end{eqnarray}
Equation (5.17) implies that $c_5$ is zero which gives that $b_z$
is zero.

\subsection{Numerical Solutions}

We consider the same circumstances as we have considered in the
rotating non-magnetized plasma and assume that
$u=\frac{1}{\sqrt{z^2+2}}=V$. Substituting this value in Eq.(5.2)
with $V_F =1$, it follows that $\lambda=1-\frac{\sqrt{z^2+2}}{z}$
which shows that the magnetic field diverges near the horizon (as
discussed in [25]). For the sake of convenience, we take
$B=\sqrt{\frac{176}{7}}$ so that $\frac{B^2}{4\pi}=2$. Making use
of $c_5=0$ (Eq.(5.17)), we get a $4\times4$ matrix of the
coefficients of constants. The complex dispersion relation has two
parts. The real part gives a dispersion relation of the type
$A(z)k^4+B(z)k^3\omega+C(z)\omega^2+D(z)\omega^4+E(z)k^2
+F(z)k^2\omega^2+G(z)k\omega+H(z)k\omega^2+L(z)=0$ whereas the
imaginary part gives a relation of the form
$A(z)k^3+B(z)k^2\omega+C(z)\omega+D(z)\omega^3+E(z)k\omega^2
+F(z)k=0.$ The first equation gives four dispersion relations out
of which two are real and interesting while the other two roots
are complex conjugates of each other. The second equation gives
three dispersion relations. The respective wave numbers are shown
in the Figures 9-11. These relations give different types of modes
when $\textbf{B}>0$ and wave number is in arbitrary direction to
$\textbf{B}.$ The two dispersion relations obtained from the real
part are shown in the Figures 7 and 8.
\begin{figure}
\center \epsfig{file=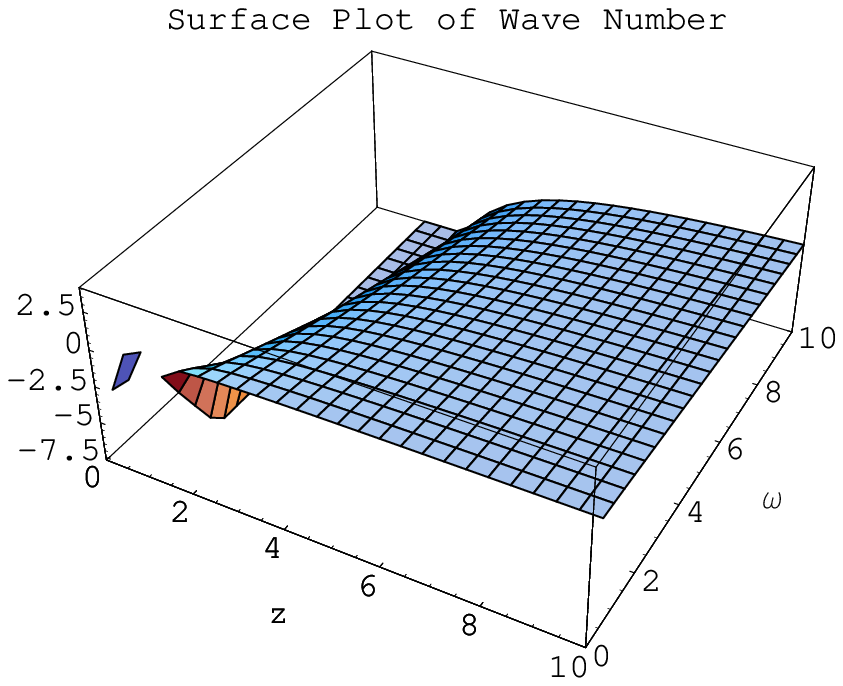,width=0.4\linewidth} \center
\epsfig{file=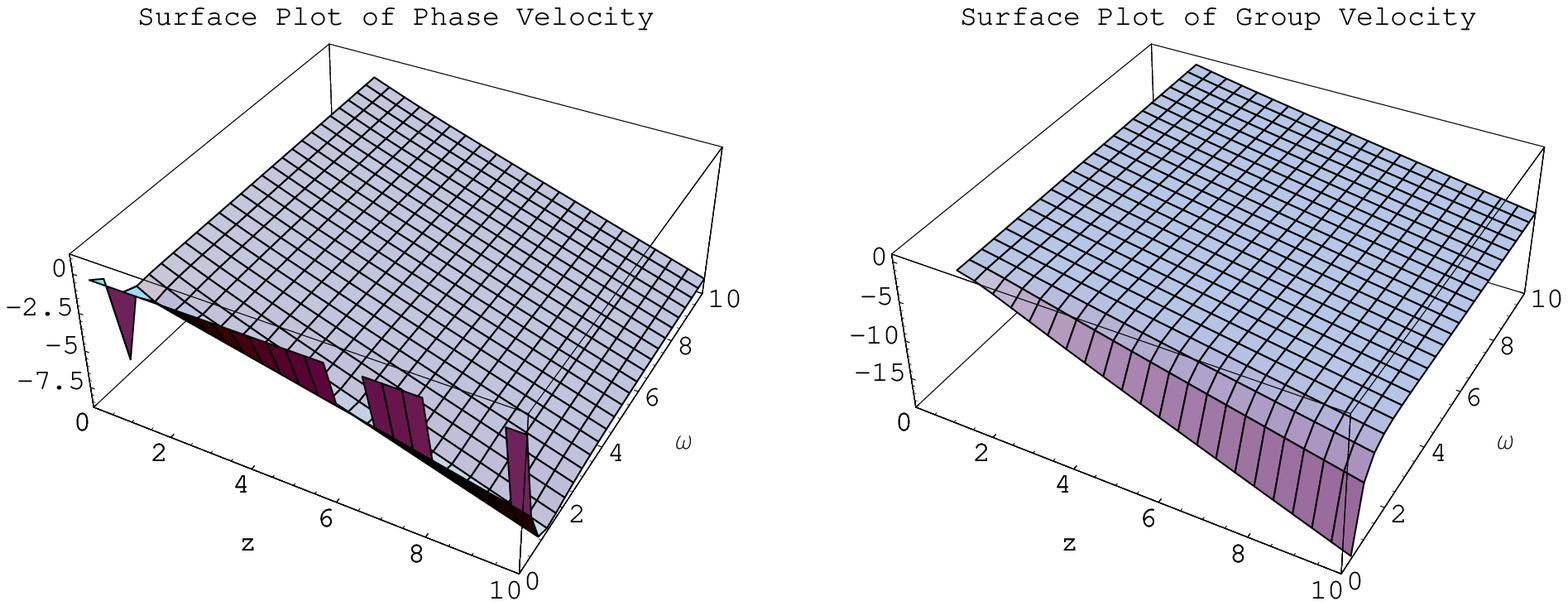,width=0.8\linewidth} \center
\epsfig{file=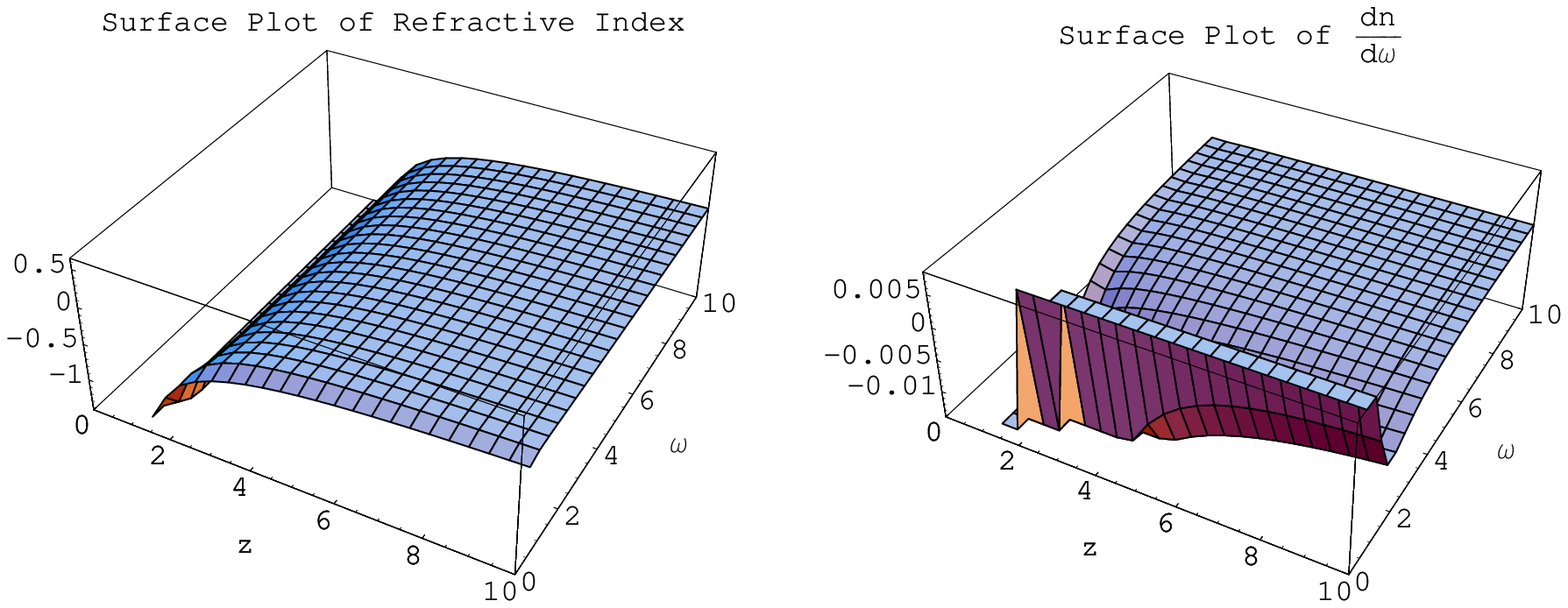,width=0.8\linewidth} \caption{The waves grow
as they move away from the event horizon. The medium admits the
properties of metamaterials.}
\end{figure}

The Figure 7 shows that at the horizon the wave number is infinite
and hence no wave is present there. $k$ has real values in the
region $1\leq z\leq10,~10^{-7}\leq\omega\leq10$. No wave is
present in the region $0\leq z<1$, evanescent waves are present in
$0\leq\omega<10^{-7}.$ All the quantities, the wave number, wave
velocity, group velocity and refractive index are negative. This
leads to the existence of metamaterial in the region [32]. The
wave number is decreasing with an increase in angular frequency
and increasing with an increase in time lapse.
\begin{figure}
\center \epsfig{file=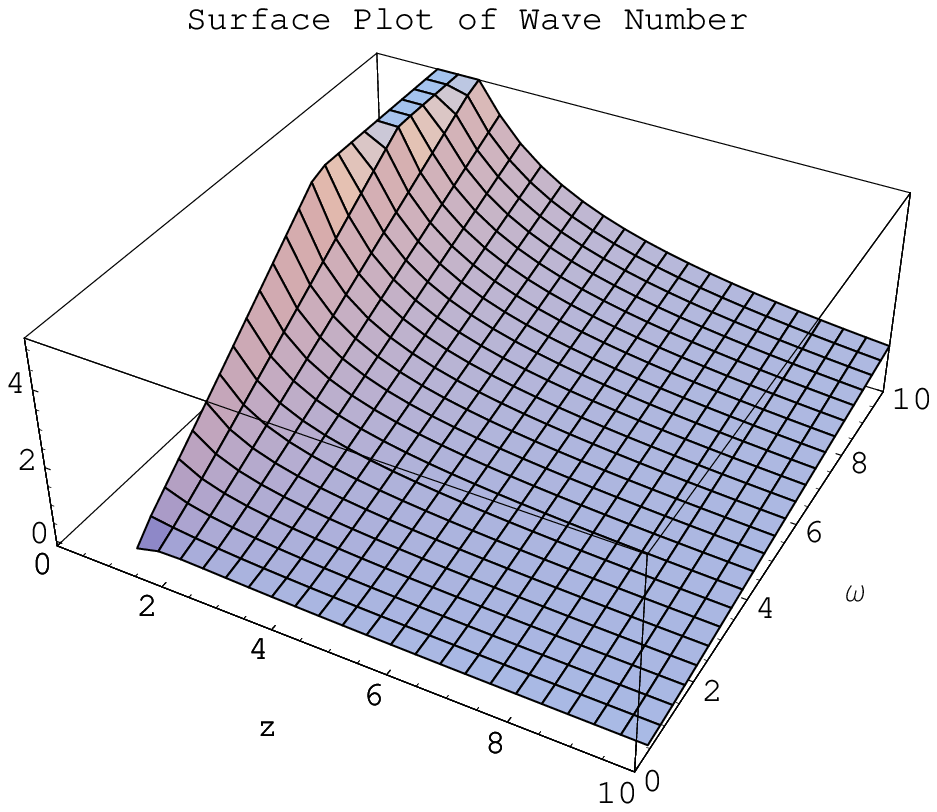,width=0.4\linewidth} \center
\epsfig{file=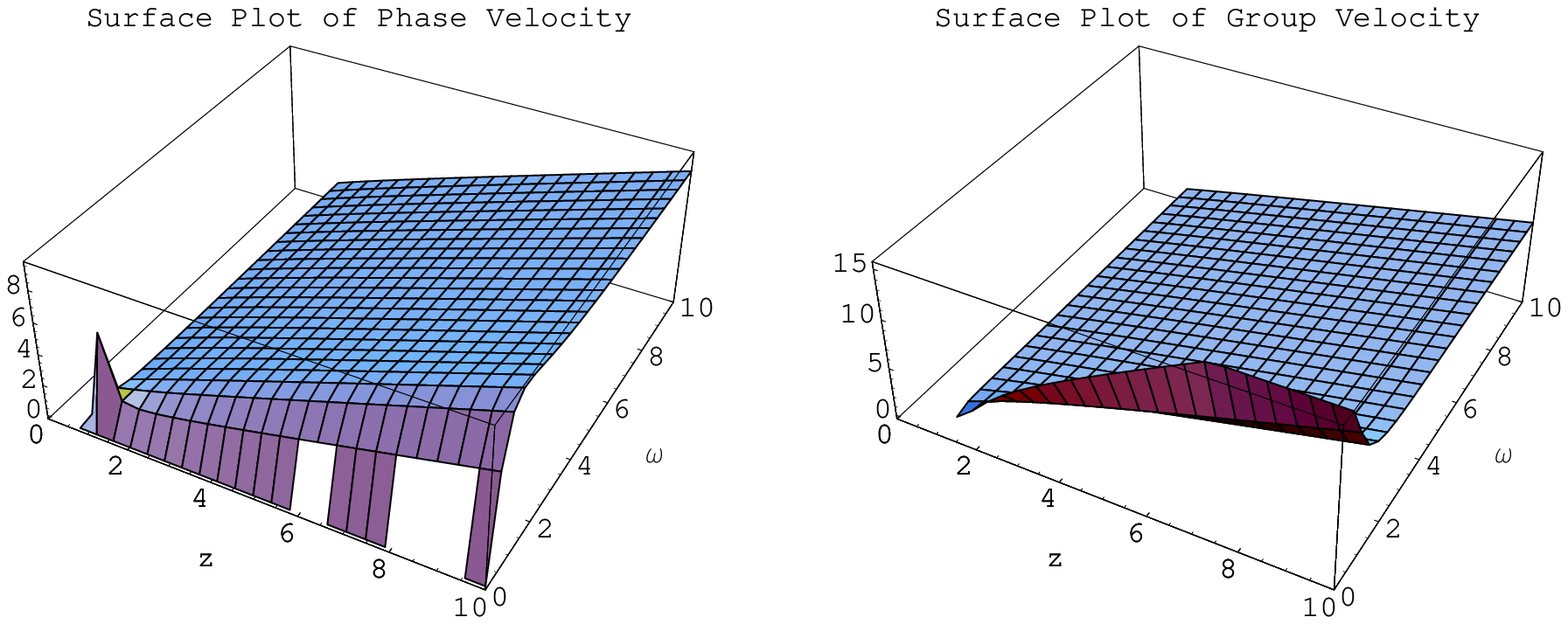,width=0.8\linewidth} \center
\epsfig{file=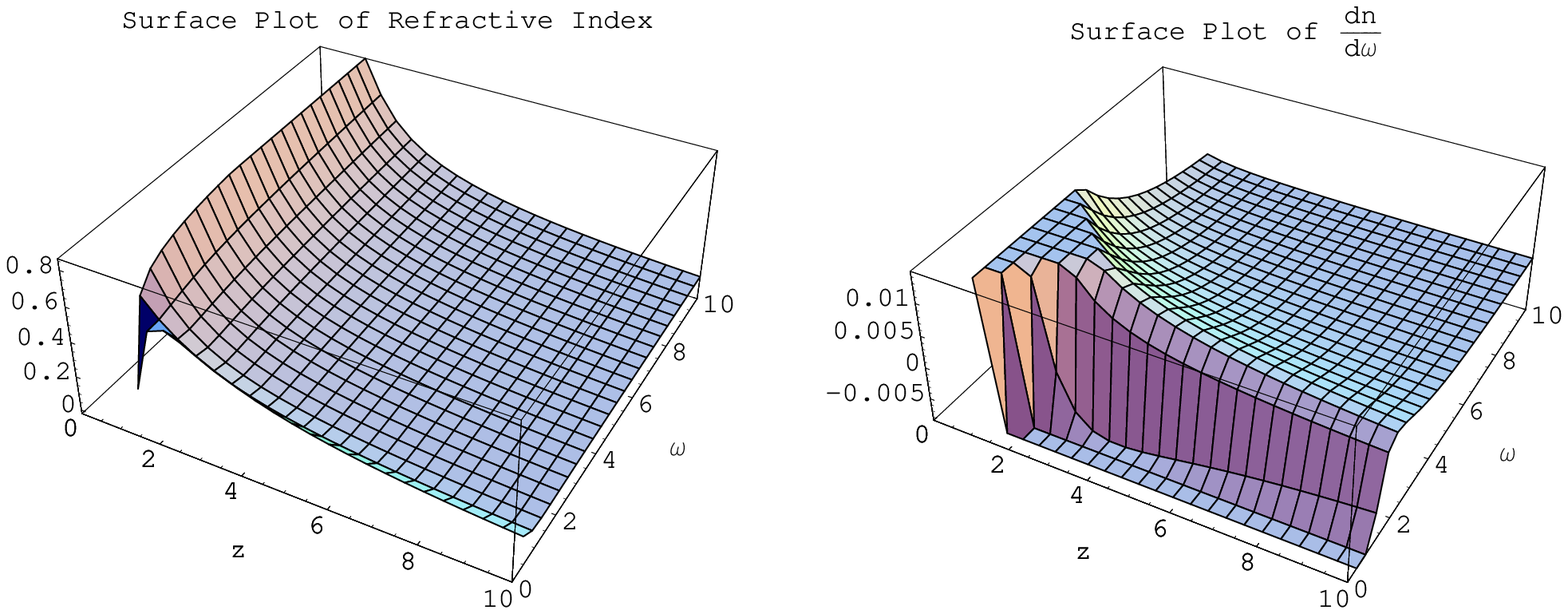,width=0.8\linewidth} \caption{The waves damp as
they move away from the event horizon. The dispersion is not normal
in the region.}
\end{figure}

The Figure 8 shows that the wave number increases with the
increase in angular velocity and decreases with the increase of
lapse function. The horizon possesses no wave due to infinite wave
number. The region $0<z\leq1$ has no wave and $0\leq\omega<5\times
10^{-6}$ is region of evanescent waves. The value of refractive
index is less than one and hence it is not a normal dispersion.

The dispersion relation obtained from the imaginary part of the
matrix determinant are shown in the Figures 9, 10 and 11.
\begin{figure}
\center \epsfig{file=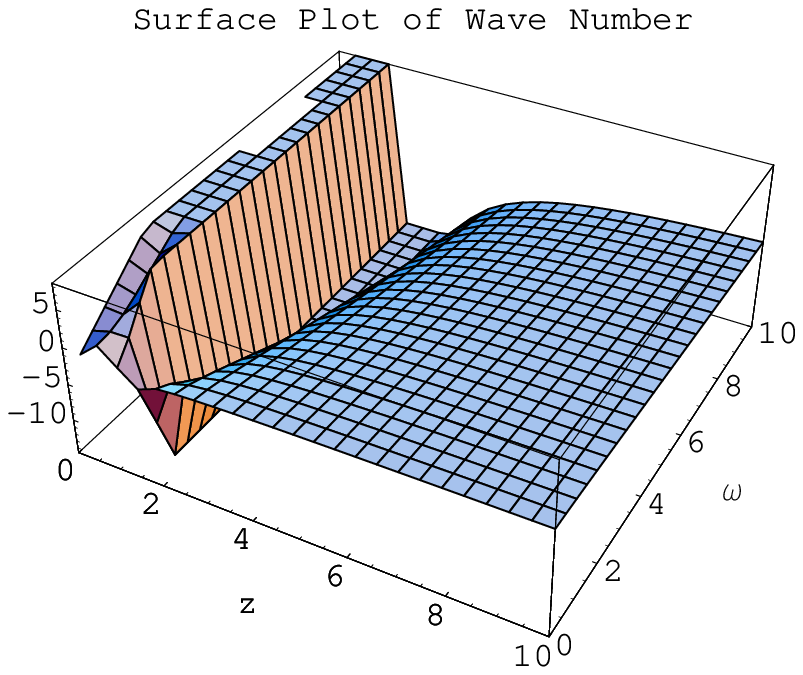,width=0.4\linewidth} \center
\epsfig{file=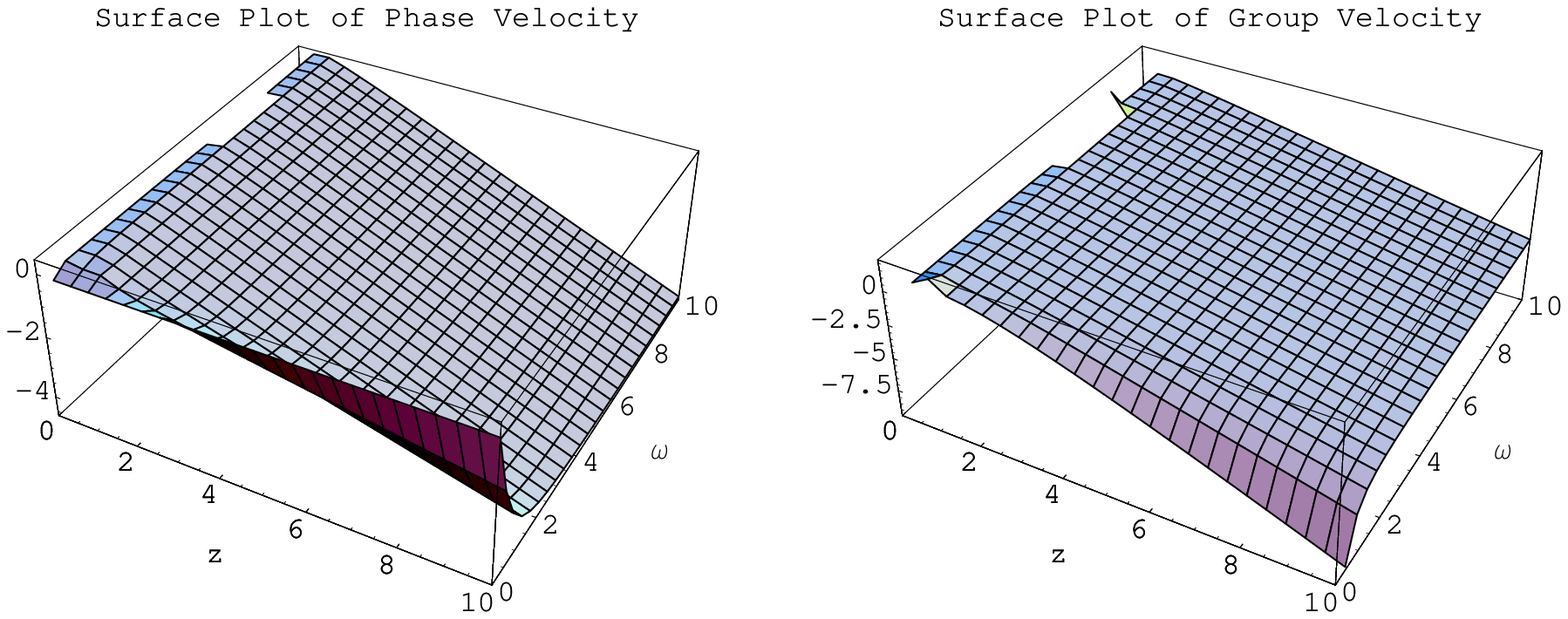,width=0.8\linewidth} \center
\epsfig{file=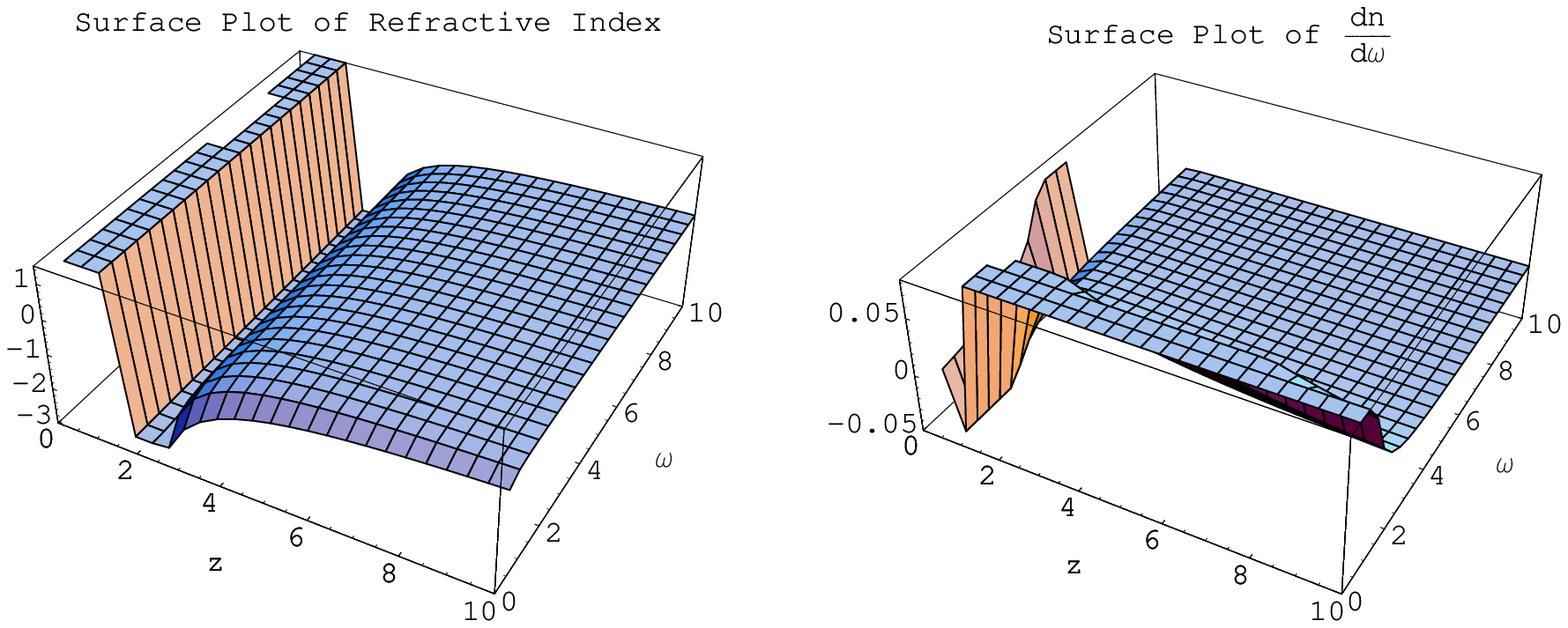,width=0.8\linewidth} \center \caption{The
waves damp and then grow as they move away from the event horizon.
The medium admits the properties of metamaterials.}
\end{figure}

The Figure 9 shows that the wave number is infinite in $0\leq
z<0.5.$ The wave number is positive in the region $0.5\leq z<1.6$,
it decreases abruptly and then it smoothly rises but is negative.
The phase and group velocities are negative. The refractive index
is negative in the domain $1.6\leq z<10$ so the region possesses
the properties of metamaterials.
\begin{figure}
\center \epsfig{file=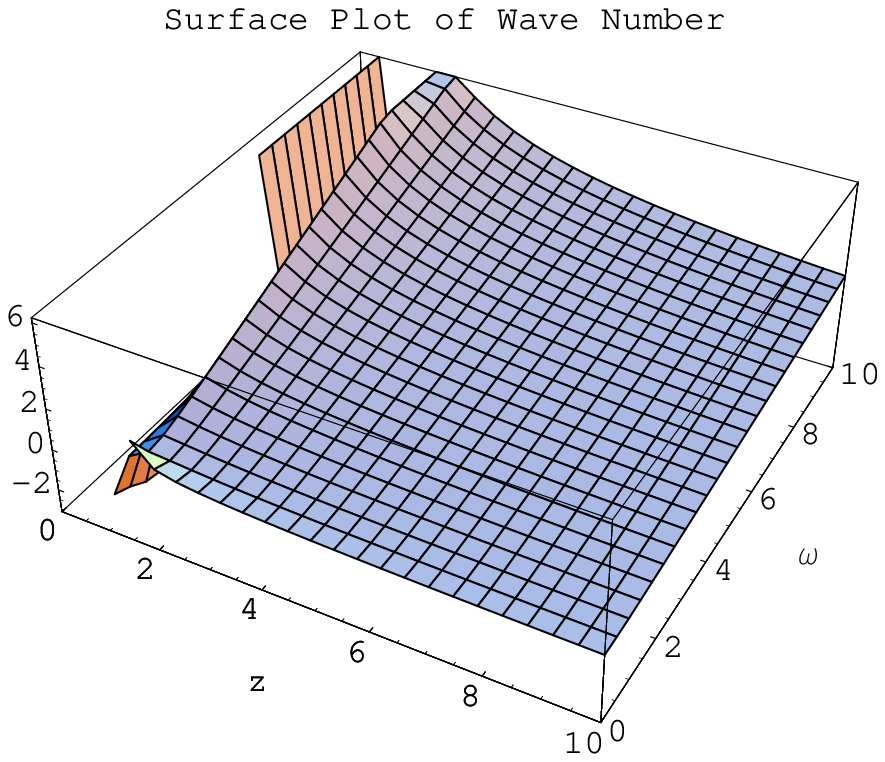,width=0.4\linewidth} \center
\epsfig{file=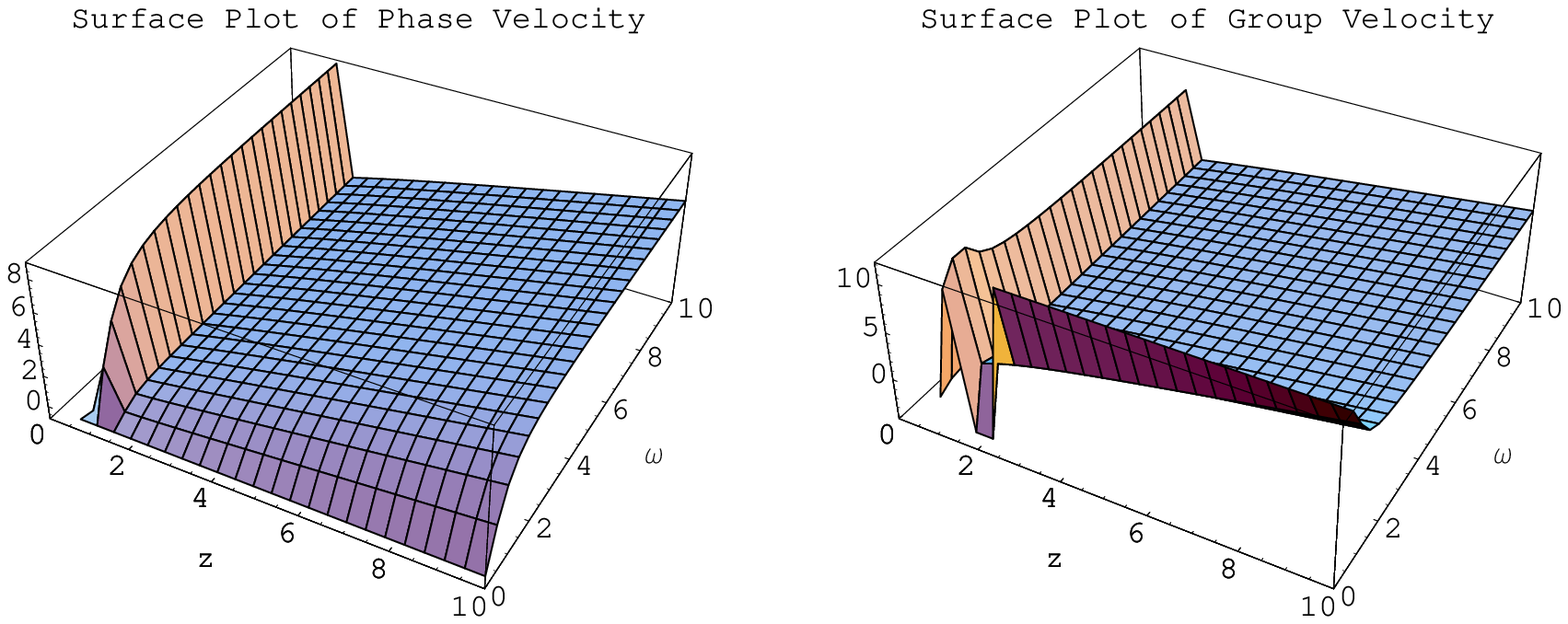,width=0.8\linewidth} \center
\epsfig{file=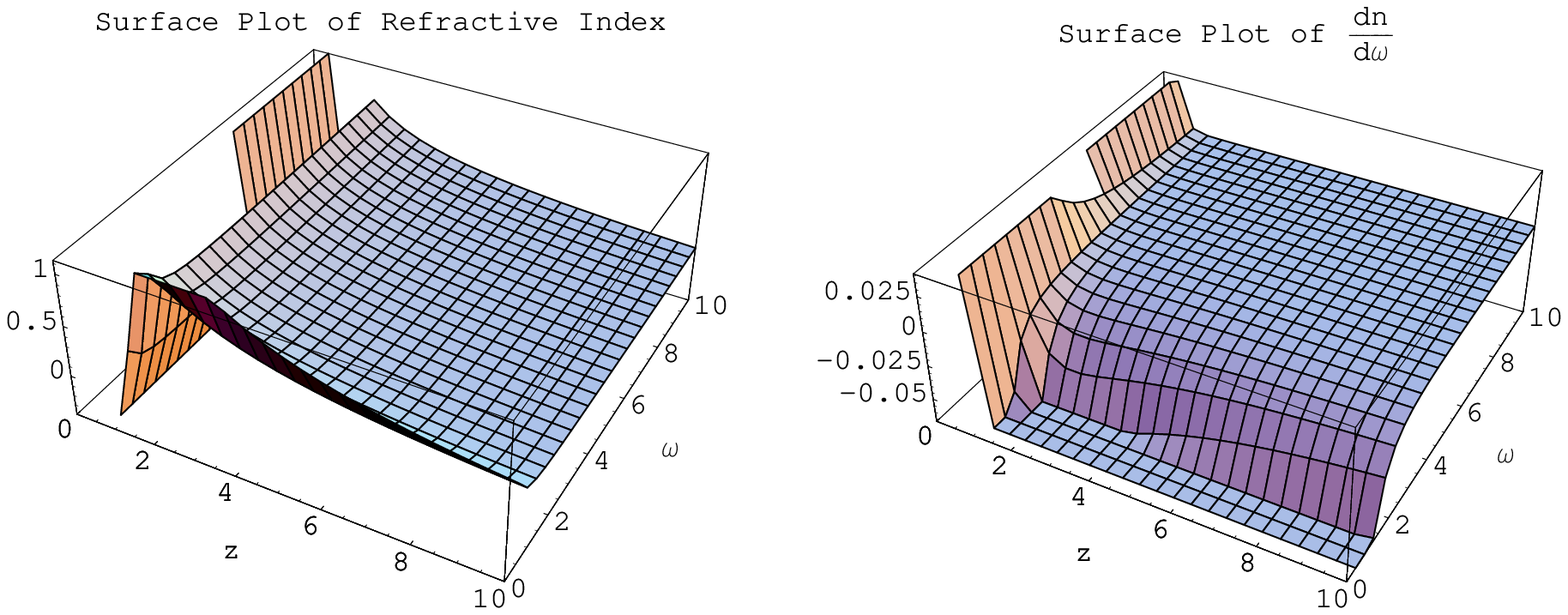,width=0.8\linewidth} \center \caption{The
waves suddenly grow and then damp on going away from the event
horizon. The dispersion is not normal.}
\end{figure}

In the Figure 10, real waves exist in the region $0.5\leq z\leq
10.$ No wave is present on the event horizon as well as in the
region $0<z\leq0.5$ because the wave number goes to infinity
there. The waves are increasing in $0.5\leq z<0.7$ then there is a
sudden decrease. The region $0.5\leq z\leq 0.7,~0\leq\omega<4$
contains evanescent waves. The refractive index is less than one
in the whole domain, hence the waves are not normally disperse.

\begin{figure} \center
\epsfig{file=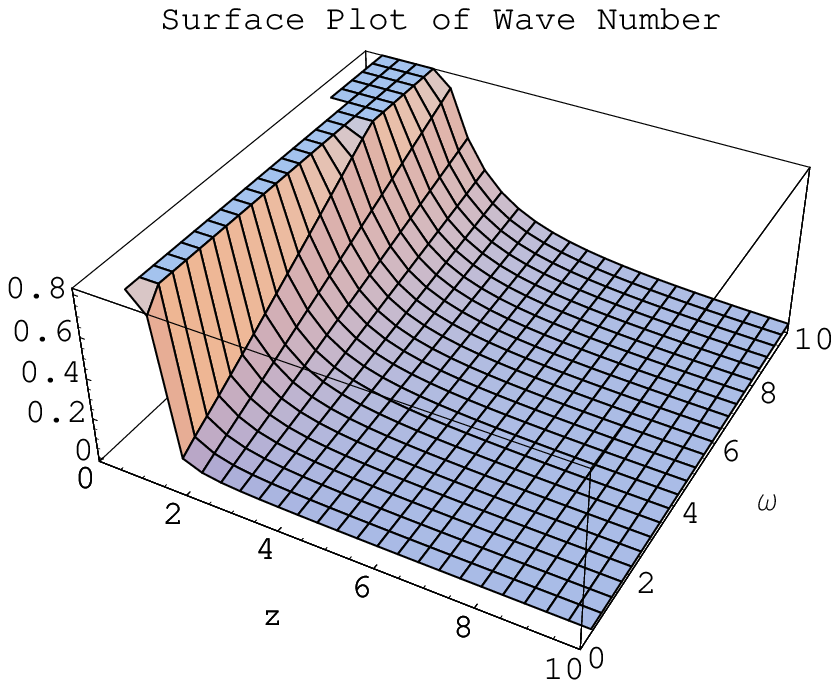,width=0.4\linewidth} \center
\epsfig{file=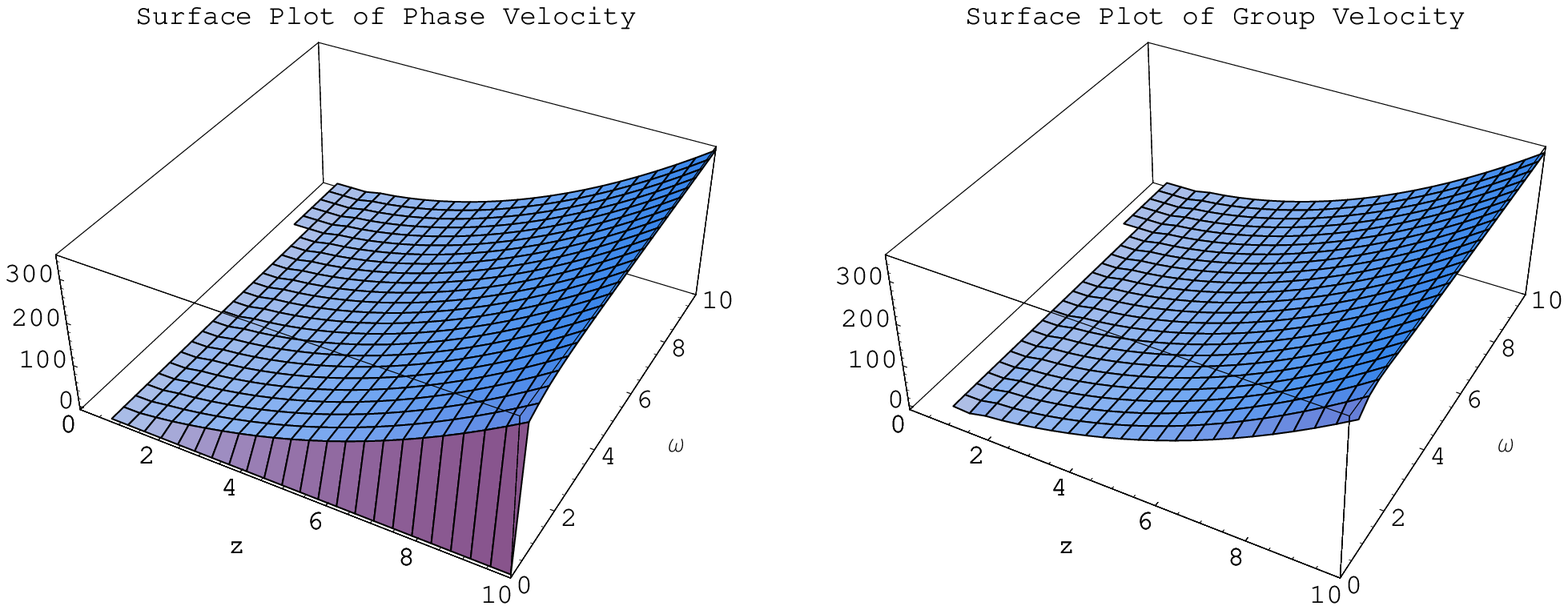,width=0.8\linewidth} \center
\epsfig{file=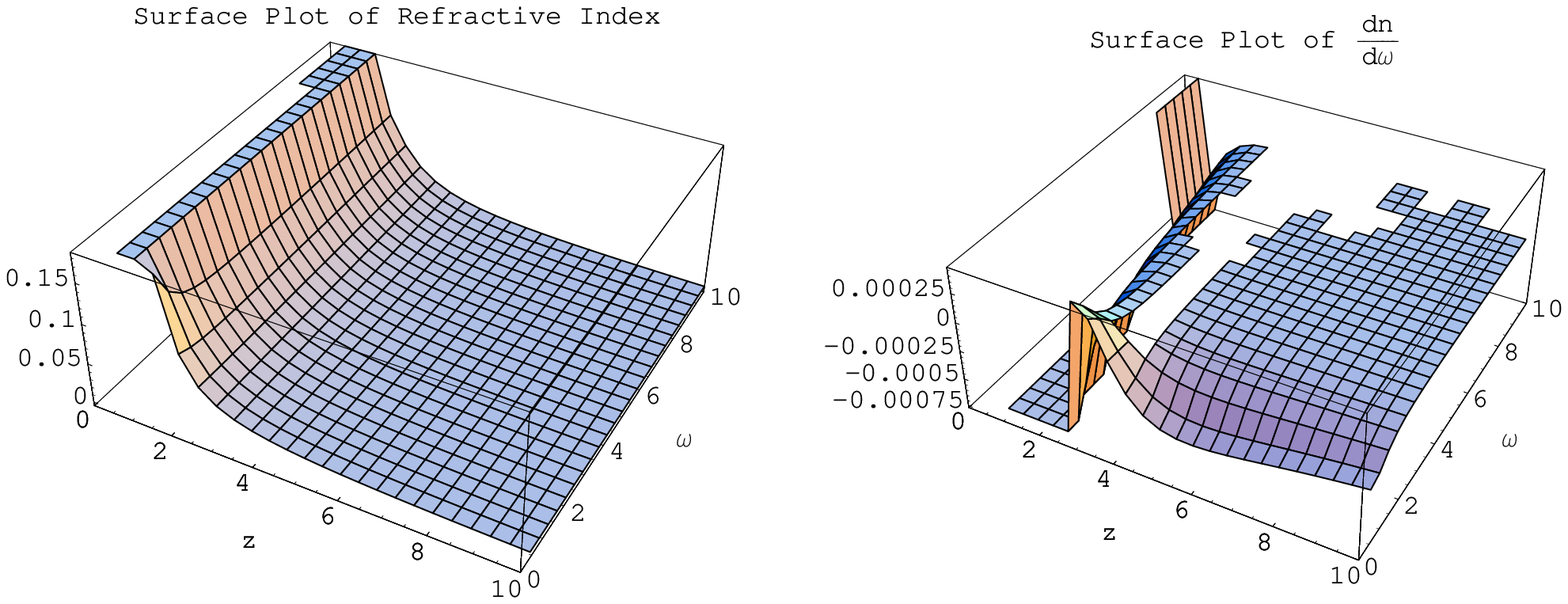,width=0.8\linewidth} \caption{The waves damp
when they move away from the event horizon. The phase and the group
velocities are equal. The dispersion is not normal.}
\end{figure}

The Figure 11 shows that real waves exist in the region $0.5\leq
z\leq 10,~0.5\leq\omega\leq10.$ The phase and group velocities are
same in this domain. The refractive index is less than one and the
dispersion is not normal. No wave is present on the event horizon
and the region near the horizon because the wave number is
infinite there. As we go away from the horizon, the wave number
decreases. On increasing the angular frequency, we find growing
waves.

\section{Conclusion\label{Conclusion}}

We have reformulated the GRMHD equations for the Schwarzschild
black hole magnetosphere to take account of gravitational effects
due to event horizon. We discuss one dimensional perturbations in
perfect MHD condition. This has been explored for the cold plasma
case. These equations are written in component form to make them
easier for the Fourier analysis. The equations are Fourier
analyzed by using the assumption of plane wave. The determinant of
the coefficients is solved numerically for the following
backgrounds in the magnetosphere.
\par\noindent
\begin{enumerate}
\item Non-Rotating Background (either non-magnetized or magnetized)
\item Rotating Non-Magnetized Background
\item Rotating Magnetized Background.
\end{enumerate}
\par\noindent
The non-rotating background shows pure Schwarzschild geometry
outside the event horizon whereas the rotating background
demonstrates the restricted Kerr geometry in the vicinity of the
event horizon which admits a variable lapse function with negligible
rotation. We have taken Schwarzschild black hole with rotating
background due to the well-known fact that the Schwarzschild black
hole is the minimum configuration of the Kerr black hole.

In the case of non-rotating background, we have found that the
magnetospheric fluid is normally disperse. The wave number becomes
indefinite at the horizon and for the values of $z$ near the
horizon and hence no wave is present there. This shows that no
signal can pass the event horizon or near to it. The third subcase
shows a mode which is not normally dispersive.

In the case of rotating non-magnetized background we have found
one case (Figure 4) in which the dispersion is normal. The regions
with zero angular frequency or very less angular frequency are
evanescent in all the three cases. The wave number becomes
infinite at event horizon and in the neighborhood of the event
horizon and no wave exists there.

In rotating magnetized background, we obtain region with negative
wave number, phase velocity, group velocity and refractive index
(Figures 7 and 9). This fact is deduced by the analysis of two of
the dispersion relations obtained in section 5. The refractive
index is negative in both the cases which implies that the region
possesses all the qualities of left-handed metamaterials. The
other three dispersion relations give that the region under
discussion is of non-normal dispersion which restrict the real
signals pass through that region.

It is worth mentioning that the wave number becomes infinite at
the event horizon and no wave exists there. This supports the
well-known point of view that no information can be extracted from
a black hole. It is interesting to mention that this work extends
the results given by Mackay et al. in [33] according to which
rotation of a black hole is required for negative phase velocity
propagation. It is also observed that the waves of less angular
velocity are evanescent. Our numerical results indicate that
negative phase velocity propagates in the rotating background
whether the black hole is rotating or non-rotating.

We know that the MHD waves in cold plasma are non-dispersive.
However, the dispersion is noted in the above figures. This factor
comes due to the formalism used and the equations which provides
different equations from the usual MHD equations. Since the 3+1
split of GR is used in a preliminary investigation of waves
propagating in a plasma influenced by the gravitational field. Thus
internal gravity waves which interrupt the MHD waves imply the cases
of dispersion in each of the hypersurface. We would like to mention
that graphs of the waves are given in a particular hypersurface
where time is constant, not in the whole Schwarzschild background
and consequently this is justified locally, not globally. Finally,
it is mentioned that some of the Figures have patches missing which
is due to the existence of complex numbers there. Mathematica cannot
plot complex numbers with real numbers. These complex numbers are
shown by gaps in the Figures.

This analysis has been done for the cold plasma state. It would be
interesting to extend this analysis for the isothermal state of
plasma. This would provide the effects of pressure in the current
work. Currently, it is in progress.

\newpage
{\bf Acknowledgment}

\vspace{0.5cm}

We acknowledge the enabling role of the Higher Education
Commission Islamabad, Pakistan, and appreciate its financial
support through the {\it Indigenous PhD 5000 Fellowship Program
Batch-II}.

\end{document}